\documentclass[aps,prd,preprint,superscriptaddress,amsmath,amssymb,showpacs]{revtex4-1}
\usepackage{dcolumn}
\usepackage{graphicx}
\usepackage{float}
\usepackage{physics}
\usepackage[colorlinks=true,allcolors=blue]{hyperref}

\usepackage{xcolor}

\begin{document}
\title{The effect of gluon condensate on the entanglement entropy in a holographic model}
\author{Xun Chen}
\email{chenxunhep@qq.com}
\affiliation{School of Nuclear Science and Technology, University of South China, Hengyang 421001, China}
\affiliation{Key Laboratory of Quark and Lepton Physics (MOE), Central China Normal University, Wuhan 430079,China.}
\author{Bo Yu}
\affiliation{School of Nuclear Science and Technology, University of South China, Hengyang 421001, China}
\author{Peng-Cheng Chu}
\email{kyois@126.com}
\affiliation{The Research Center for Theoretical Physics, Science School, Qingdao University of Technology, Qingdao 266033, China}
\author{Xiao-Hua Li}
\email{lixiaohuaphysics@126.com}
\affiliation{School of Nuclear Science and Technology, University of South China, Hengyang 421001, China}
\author{Mitsutoshi Fujita}
\email{fujitamitsutoshi@usc.edu.cn}
\affiliation{School of Nuclear Science and Technology, University of South China, Hengyang 421001, China}

\begin{abstract}
In this study, we examine the impact of the gluon condensate on holographic entanglement entropy within an Einstein-Dilaton model at both zero and finite temperatures. A critical length exists for the difference in entanglement entropy between connected and disconnected surfaces in this model, which is typically interpreted as an indicator of phase transition. As the gluon condensate increases, the critical length decreases, suggesting that confinement strengthens at zero temperature. Additionally, the entropic C-function abruptly drops to zero at the critical length, indicating the absence of entangled states. At finite temperatures, the results show that the effect of the gluon condensate on the critical length is qualitatively similar to that at zero temperature. We observe that the entropic C-function increases as a function of $L$ at finite temperature, though it exhibits competitive behaviors when the gluon condensate is large.
\end{abstract}

\maketitle
\section{Introduction}\label{sec:intro}
Quantum Chromodynamics (QCD) phase transition is a prominent research frontier in nuclear physics. The chiral symmetry in QCD is broken for finite quark masses and is gradually restored as temperature increases, leading to what is known as the chiral phase transition. Conversely, quarks are confined within hadrons at low temperatures, but at high temperatures, a quark-gluon plasma (QGP) forms, characterized by deconfined quarks from hadronic matter. It is believed that the condensation of extended structures, such as gluon rings or vortices, plays a crucial role in understanding the confinement process \cite{Houston:1979ia}. The relationship between gluon condensate and the deconfinement phase transition has been extensively studied in various papers \cite{Wang:1989ag,Baldo:2003id,Brown:2006vn,Castorina:2007qv}.

Given that the phase transition is considered a strongly coupled problem, the gauge/gravity duality has become a powerful tool. In recent decades, many phase structures and strongly coupled problems have been explored using holographic methods \cite{Herzog:2006ra,Colangelo:2010pe,Kim:2009wt,Evans:2011eu,DeWolfe:2010he,Li:2011hp,Evans:2012cx,Cai:2012xh,Cai:2012eh,Alho:2012mh,Yang:2015aia,Critelli:2016cvq,Li:2017tdz,Chelabi:2015gpc,Li:2016gfn,Li:2016smq,Li:2017ple,Knaute:2017opk,Gursoy:2017wzz,Gursoy:2018ydr,Chen:2018vty,Chen:2019rez,Chen:2017lsf,Cao:2020ske,Rodrigues:2020ndy,Arefeva:2020vae,Chen:2020ath,Zhu:2019ujc,Zhu:2019igg,Zhu:2021ucv,Giataganas:2017koz,Nakas:2020hyo,Abt:2019tas}. In holographic models, the dilaton field is essential for mimicking QCD properties \cite{Csaki:2006ji,Gubser:2008ny}. In this context, the gluon condensate is dual to the dilaton field on the gravity side. The concept of gluon condensate was introduced in \cite{Shifman:1978bx} as a measure of non-perturbative physics in QCD at zero temperature and has been widely studied at finite temperatures \cite{Lee:1989qj,DElia:2002hkf,Miller:2006hr,Brown:2006vn}. Moreover, lattice results show that the gluon condensate remains non-zero at high temperatures and undergoes significant changes near $T_c$ (the critical temperature of the deconfinement transition), regardless of the number of quark flavors. The dual geometry of gluon condensate has been proposed in early works \cite{Nojiri:1998yx,Gubser:1999pk,Kehagias:1999tr}. In recent years, the effects of gluon condensate on meson spectra, heavy-quark potential, imaginary potential, entropic destruction, Schwinger effect, and energy loss have been investigated in Refs. \cite{Ko:2009jc,Kim:2008ax,Zhao:2019tjq,Zhang:2020upv,Zhang:2020noe,Zhang:2019gki,Kim:2007qk,Afonin:2020crk}.

More than a decade ago, Shinsei Ryu and Tadashi Takayanagi proposed a holographic formula for entanglement entropy \cite{Ryu:2006bv,Ryu:2006ef} (see Refs. \cite{Nishioka:2009un,Rangamani:2016dms,Chen:2019lcd} for a review). One of the applications of holographic entanglement entropy is to detect the confinement/deconfinement transition in gauge theories \cite{Nishioka:2006gr,Klebanov:2007ws,Bah:2007kcs,Bah:2008cj,Ali-Akbari:2017vtb,Anber:2018ohz,Jain:2022hxl,daRocha:2021xwq,Zhang:2019zbf,Rahimi:2016bbv,Baggioli:2018afg,Baggioli:2020cld,Baggioli:2023ynu}. More recently, holographic entanglement has been applied to investigate the properties of the critical endpoint (CEP) and QCD phase transition \cite{Dudal:2018ztm,Mahapatra:2019uql,Arefeva:2019dvl,Knaute:2017lll,Li:2020pgn,Arefeva:2020uec,Asadi:2022mvo,Jokela:2020wgs,Fujita:2020qvp,Jain:2022hxl,Li:2024lrh}. In \cite{Dudal:2018ztm,Mahapatra:2019uql}, the phase transition between two different connected surfaces was identified as a confinement-like phase (small black hole phase) at finite temperatures in the dual QCD model. The entropic C-function in this model decreases under RG flow and exhibits a sharp drop, becoming almost zero over large intervals, which is similar to the results observed in confining theories \cite{Nishioka:2006gr,Buividovich:2008kq,Itou:2015cyu}.

In this paper, we primarily explore the relationships among the deconfinement phase transition, gluon condensate, and entanglement entropy within a holographic framework. Utilizing the dilaton black hole solution, which provides an analytic description of the interaction between a hot quark-gluon plasma and a gluon condensate without finite baryon density, we demonstrate how the gluon condensate and temperature influence the holographic entanglement entropy. Our focus is on determining whether the gluon condensate reduces the degrees of freedom in the entangled state. Notably, the holographic entanglement entropy exhibits a phase transition between connected and disconnected surfaces even at finite temperature, distinguishing our findings from those in \cite{Dudal:2018ztm,Mahapatra:2019uql}. The entropic C-function, defined as a logarithmic derivative of entanglement entropy \cite{Nishioka:2006gr}, proves to be a valuable tool for analyzing finite-temperature physics. We contend that the entropic C-function effectively captures thermal excitations at finite temperatures.

The structure of the paper is as follows: In Sec. 2, we provide a review of the Einstein-Dilaton model. In Sec. 3, we discuss the impact of the gluon condensate on holographic entanglement entropy and phase transitions at zero temperature, introducing the generalized entropic C-function to examine degrees of freedom. Sec. 4 investigates the influence of the gluon condensate on holographic entanglement entropy, the entropic C-function, and phase transitions at finite temperature. Finally, we present a summary in Sec. 5.

\vspace{10pt}
\section{A short review of the model}\label{sec:02}
We start with the Einstein-Dilaton action~\cite{Kim:2007qk,Csaki:2006ji,Ko:2009jc,Kim:2008ax,Zhao:2019tjq}
\begin{equation}
S = \frac{1}{2 \kappa^{2}} \int d^{5}x \sqrt{-G} \left( -\mathcal{R} + 2 \Lambda + \frac{1}{2} \partial_{M} \phi \partial^{M} \phi \right),
\end{equation}
where $\kappa^2$ is the five-dimensional Newton constant, $\Lambda$ is a negative cosmological constant ($\Lambda = -\frac{6}{R^2}$), and $R$ is the curvature radius. The metric is in the Einstein frame, and the Lagrangian density includes the standard Hilbert term $\sqrt{-G}\mathcal{R}/(2\kappa^2)$. Unlike typical holographic models with a dilaton potential, our toy model considers the back reaction of the gluon on the background metric, allowing us to obtain qualitative results for QCD. The Einstein equations and the equations of motion (EoM) for the scalar field are
{\begin{equation}
\begin{aligned}
\mathcal{R}_{M N}-\frac{1}{2} G_{M N} \mathcal{R}+G_{M N} \Lambda &=\frac{1}{2}\left[\partial_{M} \phi \partial_{N} \phi-\frac{1}{2} G_{M N} \partial_{P} \phi \partial^{P} \phi\right], \\
0 &=\frac{1}{\sqrt{G}} \partial_{M} \sqrt{G} G^{M N} \partial_{N} \phi.
\end{aligned}
\end{equation}}
There are two solutions to the EoM. The first is a dilaton wall solution, which deforms AdS spacetime and corresponds to a confining phase with gluon condensate at zero temperature. The second is a dilaton black hole solution, a deformation of a Schwarzschild-type AdS black hole with a dilaton background, describing gluon condensation at high temperature. The metric of the dilaton wall solution is given by:
\begin{equation}
\begin{aligned}
d s^{2} &=\frac{R^{2}}{z^{2}}\left(\sqrt{1-c^{2} z^{8}} \delta_{\mu \nu} d x^{\mu} d x^{\nu} + d z^{2}\right), \\
\phi(z)  &=\phi_{0} + \sqrt{\frac{3}{2}} \log \left(\frac{1+c z^{4}}{1-c z^{4}}\right).
\end{aligned}
\end{equation}
Here, $\phi_{0}$ and $c$ are integration constants, while $z$ denotes the radial direction. Near the UV boundary, the perturbative expansion of the dilaton field is given by
\begin{equation}
\phi(z) = \phi_{0} + \sqrt{6} c z^{4} + \ldots.
\end{equation}
We define $c = \frac{1}{z_{c}^{4}}$, where $z_{c}$ acts as an IR cutoff. The value of $c$ can be determined by the mass of the lightest glueball or heavy quarkonium~\cite{Kim:2008ax}. According to the AdS/CFT dictionary, the solution of the classical equation of motion for a scalar field $\phi$ corresponding to an operator $\mathcal{O}$ with dimension $\Delta$ has the following form near the 4D boundary as $z \rightarrow 0$,
\begin{equation}
\phi(x, z) \rightarrow z^{4-\Delta}\left[\phi_0(x) + \mathcal{O}\left(z^2\right)\right] + z^{\Delta}\left[\frac{\langle \mathcal{O}(x) \rangle}{2\Delta - 4} + \mathcal{O}\left(z^2\right)\right],
\end{equation}
where $\phi_0(x)$ acts as the source for $\mathcal{O}(x)$ and $\langle \mathcal{O}(x) \rangle$ denotes the corresponding condensate~\cite{Klebanov:1999tb, Cherman:2008eh, Afonin:2010hn}. The constant term acts as a source for the gluon condensate operator $\operatorname{Tr} G^{2}$, and the coefficient of the normalizable mode yields the gluon condensate, as discussed in Ref.~\cite{Csaki:2006ji}:
\begin{equation}
\left\langle \operatorname{Tr} G^{2} \right\rangle = \frac{8 \sqrt{3\left(N_{c}^{2} - 1\right)}}{\pi} \frac{1}{z_{c}^{4}},
\end{equation}
where we have used $\frac{1}{\kappa^{2}} = \frac{4\left(N_{c}^{2} - 1\right)}{\pi^{2} R^{3}}$, and $N_c$ is the number of colors. For qualitative analysis in this work, we set $\phi_{0} = 0$. The next solution to consider is the dilaton black hole background. We have
\begin{equation}
d s^{2}=\frac{1}{z^{2}}\left(A d \vec{x}^{2}+B d t^{2}+d z^{2}\right),
\end{equation}
where
\begin{equation}
\begin{aligned}
A &=\left(1+f z^{4}\right)^{(f+a) / 2 f}\left(1-f z^{4}\right)^{(f-a) / 2 f}, \\
B &=\left(1+f z^{4}\right)^{(f-3 a) / 2 f}\left(1-f z^{4}\right)^{(f+3 a) / 2 f}, \\
f^{2} &=a^{2}+c^{2} .
\end{aligned}
\end{equation}
We observe that this dilaton black hole solution reduces to the AdS black hole solution when $c = 0$, and it simplifies to the dilaton-wall background when $a = 0$. Following Ref.~\cite{Kim:2008ax}, the position of the IR cutoff is given by $z_{f} = f^{-1/4}$. The temperature is related to $a$ by $a = (\pi T)^4 / 4$.

\vspace{10pt}
\section{The effect of gluon condensate on holographic entanglement entropy at vanishing temperature}\label{sec:03}
\begin{figure*}
	\centering
	\includegraphics[width=15cm]{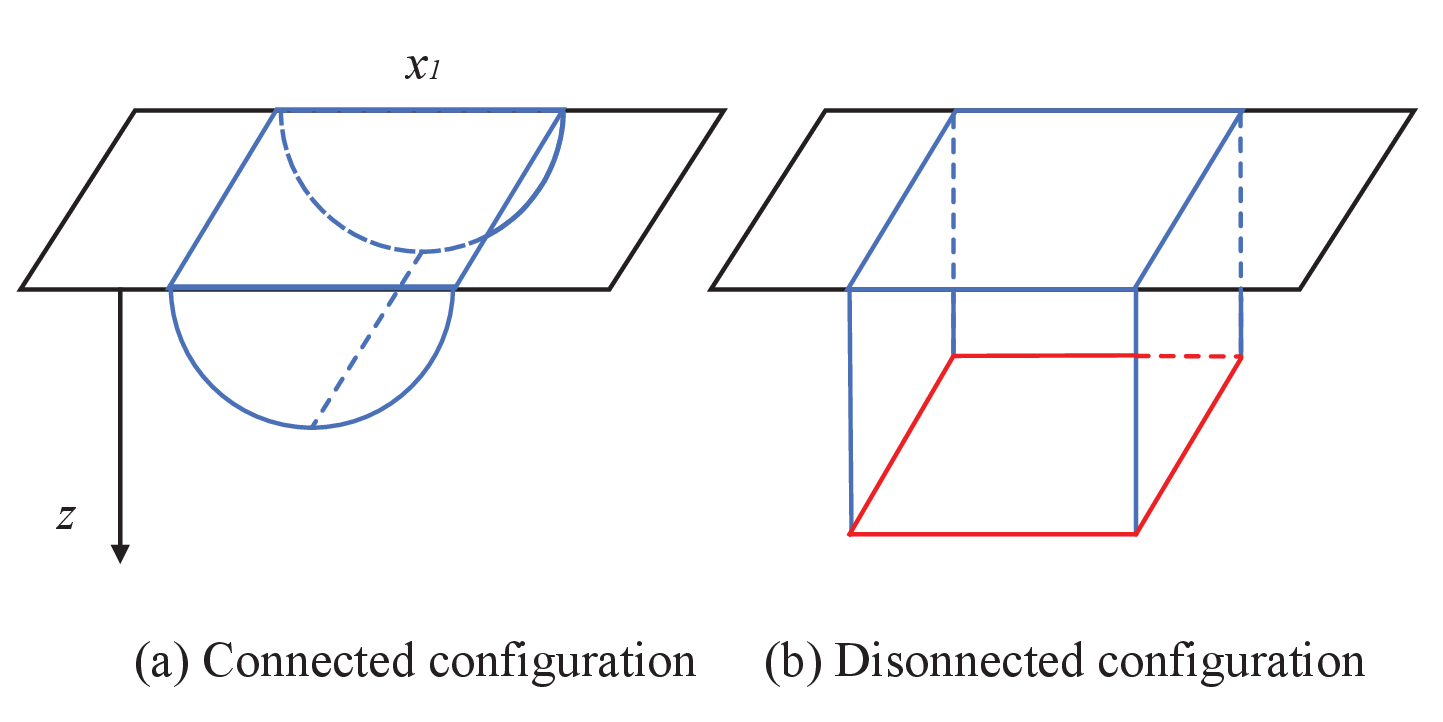}
	\caption{\label{Sketch} A schematic diagram of connected configuration and disconnected configuration. }
\end{figure*}

In this section, we explore the effect of the gluon condensate on the holographic entanglement entropy at zero temperature and present some numerical results. We consider a quantum mechanical system described by the density operator $\rho_{\text{tot}}$, partitioned into a subsystem $\mathcal{A}$ and its complement $\mathcal{B}$. The entanglement entropy of $\mathcal{A}$ is defined as the von Neumann entropy:
\begin{equation}
S_{\mathrm{EE}} := -\operatorname{Tr}_{\mathcal{A}} \rho_{\mathcal{A}} \ln \rho_{\mathcal{A}},
\end{equation}
where $\rho_{\mathcal{A}} = \operatorname{Tr}_{\mathcal{B}} \rho_{\text{tot}}$ is the reduced density matrix, and the density matrix of the pure ground state $|\Psi\rangle$ is given by $\rho_{\text{tot}} = |\Psi\rangle \langle \Psi|$.
According to the original Refs.~\cite{Ryu:2006bv, Ryu:2006ef}, the holographic dual of this quantity for a ${\rm{CFT}}_{d}$ on $\mathbb{R}^{1, d-1}$ is given by
\begin{equation}
S_{\mathrm{HEE}} = \frac{\operatorname{Area}(\gamma_{\mathcal{A}})}{4 G_{(d+1)}}.
\end{equation}
Here, $\gamma_{\mathcal{A}}$ is the static minimal surface in $\mathrm{AdS}_{d+1}$ with the boundary condition $\partial \gamma_{\mathcal{A}} = \partial \mathcal{A}$, and $G_{(d+1)}$ is the $(d+1)$-dimensional Newton constant. We assume a fixed strip shape on the boundary for the entanglement region
$$
\mathcal{A}: \quad x_{1} \in \left[-\frac{l}{2}, \frac{l}{2}\right], \quad x_{2}, x_{3} \in (-\infty, \infty).
$$
Then, the minimal area of $\gamma_{\mathcal{A}}$, which is proportional to the entanglement entropy of the subsystem $\mathcal{A}$, is obtained by minimizing the following area:
\begin{equation}\label{SA}
S_{\mathcal{A}}^{(c)} = \frac{L}{4 G_{5}} \int d^{3}x \sqrt{g_{\text{ind}}},
\end{equation}
where $g_{\text{ind}}$ is the induced metric on $\gamma_{\mathcal{A}}$. We assume a general form of the metric as
\begin{equation}\label{ds}
d s^{2} = -f_{1}(z) d t^{2} + f_{2}(z) d z^{2} + f_{3}(z) d \vec{x}^{2}.
\end{equation}
Using Eqs. (\ref{SA}) and (\ref{ds}), we can derive that
\begin{equation}\label{SAA}
S_{\mathcal{A}}^{(c)} = \frac{V_{2}}{4 G_{5}} \int_{-\frac{l}{2}}^{\frac{l}{2}} d x_1 \sqrt{f_{3}^{3}(z) + f_{3}^{2}(z) f_{2}(z) \left(z'\right)^2},
\end{equation}
where $V_{2}$ is the area of the two-dimensional surface defined by $x_{2}$ and $x_{3}$, and $z' = \frac{d z}{d x_1}$. The above area does not explicitly depend on $x_1$, so the corresponding Hamiltonian is a constant of motion:
\begin{equation}\label{hamiltonian}
\frac{f_{3}^{2}(z)}{\sqrt{f_{3}^{3}(z) + f_{2}(z) \left(z'\right)^2}} = \text{const} = f_{3}^{\frac{3}{2}}(z_{*}),
\end{equation}
where $z_{*}$ is the maximum value of $z$, i.e., $z(x_1 = 0) = z_{*}$ and $z'(x_1 = 0) = 0$. Thus, from ~(\ref{hamiltonian}), we can get
\begin{equation}\label{zp}
z^{\prime}=\sqrt{\frac{f_{3}(z)}{f_{2}(z)}} \sqrt{\frac{f_{3}^{3}(z)}{f_{3}^{3}\left(z_{*}\right)}-1}.
\end{equation}
The relation between $L$ and $z_{*}$ can be obtained as
\begin{equation}
L=2 \int_{0}^{z_{*}} \sqrt{\frac{f_{2}(z)}{f_{3}(z)}} \frac{d z}{\sqrt{\frac{f_{3}^{3}(z)}{f_{3}^{3}\left(z_{*}\right)}-1}}.
\end{equation}
At last, from ~(\ref{hamiltonian}) and ~(\ref{zp}), we can see that
\begin{equation}\label{SC}
S_{A}^{(c)}=\frac{V_{2}}{2 G_{5}} \int_{0}^{z_{*}} \frac{f_{3}^{\frac{5}{2}}(z) f_{2}^{\frac{1}{2}}(z)}{\sqrt{f_{3}^{3}(z)-f_{3}^{3}\left(z_{*}\right)}} d z.
\end{equation}
Another configuration we will consider here is a disconnected solution. This configuration is described by two disconnected surfaces located at $x_1 = L/2 $ and extended in all other spatial directions. We get the following expression for the disconnected solution
\begin{equation}
S_{A}^{(d)}=\frac{V_{2}}{2 G_{5}} \int_{0}^{z_c} f_{3}(z) f_{2}^{\frac{1}{2}}(z) d z.
\end{equation}
The connected and disconnected configuration are shown in Fig.~\ref{Sketch} and define
\begin{equation}
\Delta S(l) \equiv \frac{2 G_{5}}{V_{2}}\left(S_{A}^{(c)}-S_{A}^{(d)}\right).
\end{equation}

\begin{figure*}
	\centering
	\includegraphics[width=15cm]{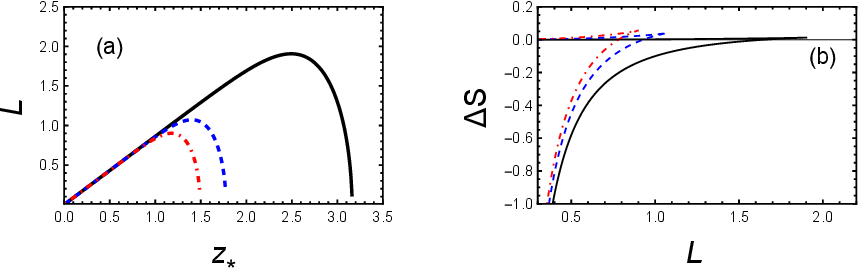}
	\caption{\label{zeroSL}(a) Strip length $L$ as a function of $z_*$ for $c = 0.01$ $\rm GeV^4$ (solid black line), $0.1$ $\rm GeV^4$ (blue dashed line), $0.2$ $\rm GeV^4$ (red dot-dashed line). (b) Difference in entanglement entropy between the connected and disconnected surface as a function of the length of the strip $L$ for $c = 0.01$ $\rm GeV^4$ (solid black line), $0.1$ $\rm GeV^4$ (blue dashed line), $0.2$ $\rm GeV^4$ (red dot-dashed line). The units of $L$ and $z_*$ are $\rm GeV^{-1}$.}
\end{figure*}

Next, we present the numerical results in Fig.~\ref{zeroSL}. The left panel of Fig.~\ref{zeroSL} illustrates that the strip length increases with $z_*$, reaching a maximum value $L_{\text{max}}$. Beyond this point, the connected entangling surface does not exist, and only the disconnected entangling surface remains. As the gluon condensate increases, the maximum strip length $L_{\text{max}}$ decreases. We also observe that $\Delta S$ changes sign at a critical length $L_c$, which is smaller than $L_{\text{max}}$. This indicates that the disconnected surface becomes dominant when $L > L_c$. Therefore, a phase transition occurs at $L = L_c$, corresponding to the confinement/deconfinement phase transition in the dual gauge theory. As the gluon condensate increases, the critical length $L_c$ decreases. This suggests that the confined phase predominates at large gluon condensate and zero temperature. Consequently, the leftward shift of the critical length $L_c$ with increasing gluon condensate implies that a large gluon condensate promotes confinement.

\begin{figure*}
	\centering
	\includegraphics[width=15cm]{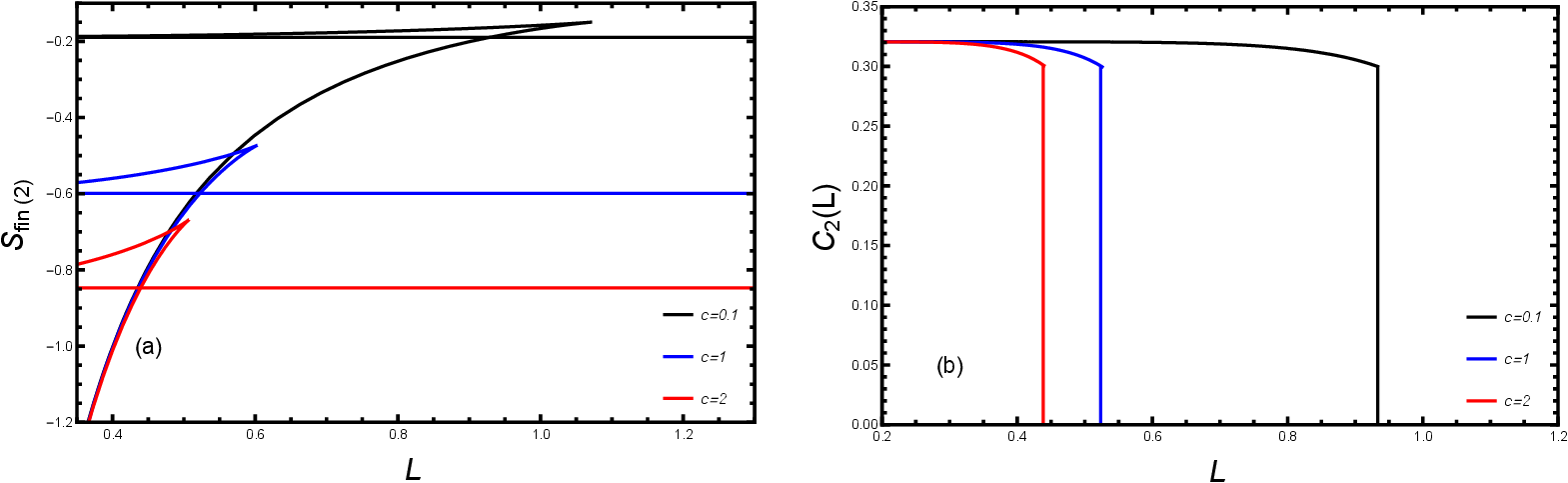}
	\caption{\label{zeroSL2} (a) The finite part of holographic entanglement entropy $S_{fin(2)}=2G_5S_{fin}/V_2$ as a function of strip length $L$. (b) Normalized entropic C-function $C_2(L)=2G_5C(L)$ as a function of $L$.
}
\end{figure*}

Moreover, we analyze confinement from different aspects. Holographic entanglement entropy of disconnected configuration becomes constant and periodic as a function of $L$. It implies that there are no entangled states (e.g. product states). Thus, one does not need to subtract this part to see confinement in detail. Instead, the finite part of holographic entanglement entropy is defined as
\begin{equation}
S_{fin}=S_A-\dfrac{V_2}{4G_5\epsilon^2}.
\end{equation}
Thus, the divergent part is subtracted. In Fig. \ref{zeroSL2} (a), normalized $S_{fin}$ is plotted as a function of $L$. One can see that this finite part of the connected surface decreases with increase of gluon condensate. Because degrees of freedom of entangled states decrease, gluon condensate contributes confinement. Increase of condensate decreases $S_{fin}$ of the connected surface and $S_{fin}$ of the disconnected surface is a constant. $C(L)$ suddenly jumps to zero at large $L$ shown in Fig. \ref{zeroSL2} (b).

Generalized entropic C-function is more relevant for confinement. It is defined as
\begin{equation}
C(L)=L^3\dfrac{\partial S_A}{V_2\partial L}.
\end{equation}
This is a generalization of $2d$ entropic C-function~\cite{Casini:2004bw,Casini:2006es}. Generalized entropic C-function represents degrees of freedom at the energy scale $E\sim 1/L$. Normalized entropic C-function is plotted as a function of $L$ in Fig. \ref{zeroSL2} (b). It decreases as $L$ increases and suddenly jumps to zero. For large $L$, degrees of freedom of entangled states are not remained. This is consistent with the analysis of meson mass in \cite{Kim:2008ax}. Since the critical length $L_c$ decreases with increase of gluon condensate, confinement is favored for large gluon condensate and even for large energy.

According to the Ref. \cite{Itou:2015cyu}, the critical temperature of the pure SU(3) Yang-Mills theory is $T_c^{-1}=0.714$ fm. In SU(2) gauge theory, the entropic C-function shows a clear discontinuity around $L=0.5$ fm. For the holographic model with gluon condensate and for $c=0.01$, the critical length is $L_c=0.2\times 1.7=0.34$ fm. It is the same order as the above-mentioned scales. Besides, the entropic C-function calculated in lattice is 0.206 for $0 \leq L \leq 0.7$ fm. In our paper, the entropic C-function for small $L$ is 0.32 which is also very close to lattice QCD. To increase the critical length, one needs to have smaller gluon condensate.

\section{The effect of gluon condensate on holographic entanglement entropy at finite temperature}\label{sec:04}
\begin{figure*}
	\centering
	\includegraphics[width=15cm]{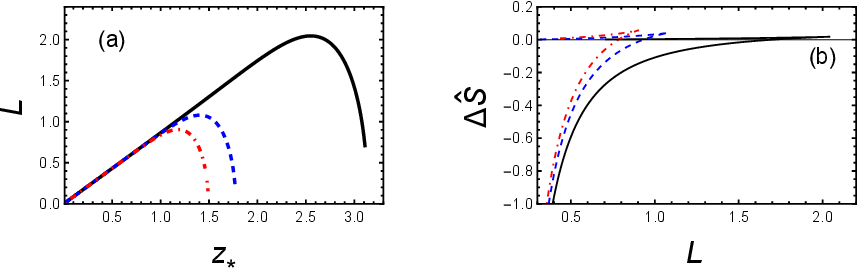}
	\caption{\label{CSL} (a) Strip length $L$ as a function of $z_*$ for $c = 0.01$ $\rm GeV^4$ (black solid line), $0.1$ $\rm GeV^4$ (blue dashed line), $0.2$ $\rm GeV^4$ (red dot-dashed line) at a fixed $T = 0.1$ $\rm GeV$. (b) The difference of entanglement entropy between the connected and disconnected surface as a function of strip length $L$ for $c = 0.01$ $\rm GeV^4$ (black solid line), $0.1$ $\rm GeV^4$ (blue dashed line), $0.2$ $\rm GeV^4$ (red dot-dashed line) at a fixed $T = 0.1$ $\rm GeV$. The units of $L$ and $z_*$ are $\rm GeV^{-1}$.}
\end{figure*}

\begin{figure*}
	\centering
	\includegraphics[width=15cm]{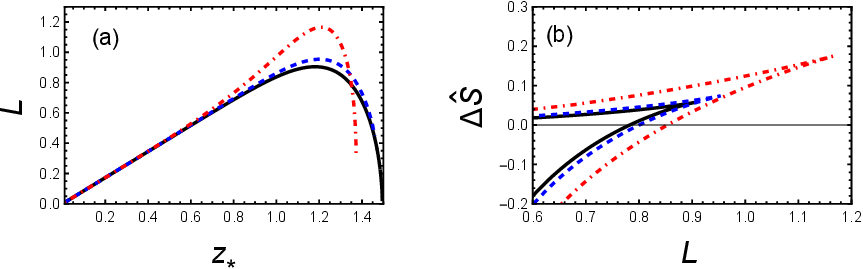}
	\caption{\label{TSL}(a) Strip length $L$ as a function of $z_*$ for $T = 0.1$ $\rm GeV$ (black solid line), $0.2$ $\rm GeV$ (blue dashed line), $0.3$ $\rm GeV$ (red dot-dashed line) at a fixed $c = 0.2$ $\rm  GeV^4$. (b) The difference of entanglement entropy between the connected and disconnected surface as a function of strip length $L$ for $T = 0.1$ $\rm GeV$ (black solid line), $0.2$ $\rm GeV$ (blue dashed line), $0.3$ $\rm GeV$ (red dot-dashed line) for a fixed $c = 0.2$ $\rm GeV^4$. The units of $L$ and $z_*$ are $\rm GeV^{-1}$.}
\end{figure*}

\begin{figure*}
	\centering
	\includegraphics[width=15cm]{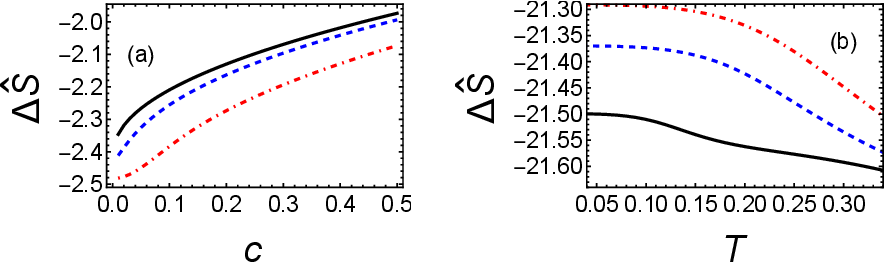}
	\caption{\label{TCS}(a) The difference of entanglement entropy $\Delta S$ as a function of gluon condensate $c$ for $T = 0.1$ $\rm GeV$ (black solid line), $0.2$ $\rm GeV$ (blue dashed line), $0.3$ $\rm GeV$ (red dot-dashed line). (b) The difference of entanglement entropy $\Delta S$ as a function of temperature $T$ for $c = 0.01$ $\rm GeV^4$ (black solid line), $0.1$ $\rm GeV^4$ (blue dashed line), $0.2$ $\rm GeV^4$ (red dot-dashed line). }
\end{figure*}
\begin{figure*}
	\centering
	\includegraphics[width=15cm]{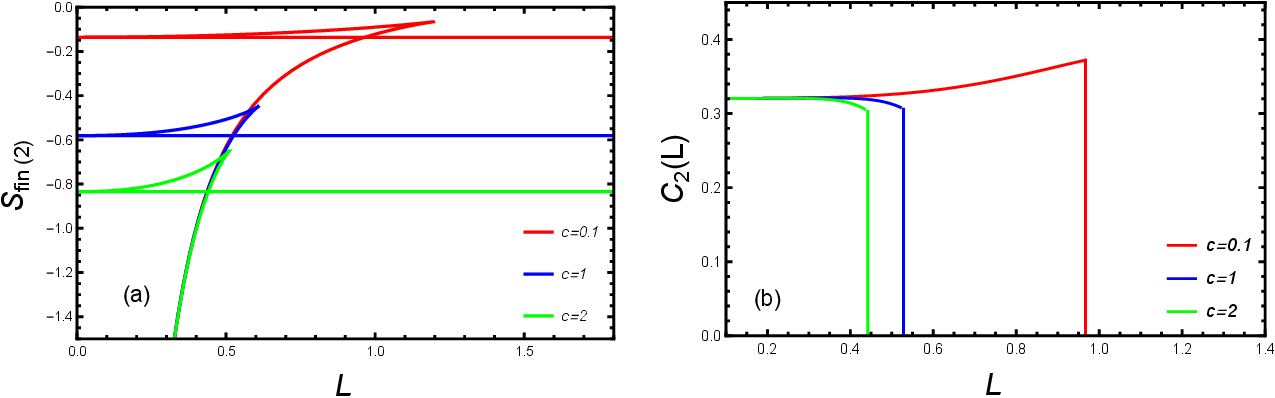}
	\caption{\label{TCS2}(a) The normalized finite part of holographic entanglement entropy as a function of $L$ for fixed temperature $T=0.2$ GeV. The finite part of the disconnected surface is a constant. (b) the normalized $C_2(L)$ as a function of $L$ for fixed temperature $T=0.2$ GeV.  }
\end{figure*}

In this section, we will turn to finite temperature and see the difference with previous case. The entanglement entropy of connected surface is the same as ~(\ref{SC}) at finite temperature. But the disconnected surface is a little bit different. Considering
\begin{equation}
x=-\frac{L}{2}, \quad z=z_{f}, \quad x=\frac{L}{2},
\end{equation}
we can get~\cite{Ali-Akbari:2017vtb,Dudal:2018ztm}
\begin{equation}
\hat{S}_{A}^{(d)}=\frac{V_{2}}{4 G_{5}}\left(2 \int_{0}^{z_{f}} d z f_{3}(z) \sqrt{f_{2}(z)}+L \sqrt{f_{3}^{3}\left(z_{f}\right)}\right).
\end{equation}
Similarly, the difference of entanglement entropy can be defined as
\begin{equation}
\Delta \hat{S} \equiv \frac{2 G_{5}}{V_{2}}\left(\hat{S}_{A}^{(c)}-\hat{S}_{A}^{(d)}\right).
\end{equation}
We show the numerical results in Fig.~\ref{CSL} and Fig.~\ref{TSL} for fixed temperature and gluon condensate, respectively. The Fig.~\ref{CSL} also shows strip length is increasing with $z_*$ for a fixed temperature. Similar as previous case, the critical length $L_c$ will shift to the left, which again implies the confined phase tends to dominate when we increase the gluon condensate. In Fig.~\ref{TSL}, we fix the value of gluon condensate and change the temperature. The qualitative behavior of temperature in this model is consistent with the Refs.~\cite{Ali-Akbari:2017vtb,Dudal:2018ztm}.

To be more clear, we show the the difference of entanglement entropy as a function of temperature and gluon condensate in Fig.~\ref{TCS}. When we increase the value of gluon condensate, the difference of entanglement entropy will increase. It means the connected configuration will dominate with the increase of gluon condensate, which is in favor of the confined phase. The critical length will become smaller at the same time. When we increase the value of temperature, the difference of entanglement entropy will decrease, which means the critical length will become larger. Deconfined phase will dominate with the increase of the temperature.

Moreover, we compute the finite part of holographic entanglement entropy as a function of $L$ in Fig.~\ref{TCS2} (a). This finite part can be considered as degrees of freedom of entangled states. The finite part of connected surfaces decreases with increase of gluon condensate in a small way, while it increases with increase of temperature (see Fig. \ref{zeroSL2} (a) for $T=0$). In Fig.~\ref{TCS2} (b), the generalized entropic C-function is plotted as a function of $L$. For $T=0.2$ GeV and $c=0.1$ $\rm GeV^4$, the entropic C-function increases (see Fig. \ref{zeroSL2} (b) for $T=0$). It captures degrees of freedom of thermal entangled state. When $c$ increases, entropic C-function decreases because of the competitive behavior between confinement and deconfinement.

The generalized entropic C-function quantifies the degrees of freedom at the energy scale of $1/L$. As depicted in Fig.~\ref{TCS2} (b), it accounts for thermal excitations of order $T$ due to the rise in the entropic C-function at finite temperatures ($TL\sim1$). Conversely, thermal excitations are diminished by a substantial gluon condensate, which results from the competitive dynamics between the two. This particular insight is not discernible from the entanglement entropy alone. The component of thermal excitations is analogous to thermal entropy, which emerges from the renormalized entanglement entropy at high temperatures in the context of Ref. \cite{Liu:2012eea}.

The generalized entropic C-function also helps the understanding of gluon condensate during confinement. The fact that confinement is favored for large gluon condensate is consistent with the analysis of lattice QCD. The gluon condensate is larger in the confining phase rather than the deconfinement phase around $T_c$ (there is a drop of the gluon condensate after the confinement/deconfinement phase transition). Besides, the entropic C-function can probe the confinement/deconfinement phase transition in the presence of gluon condensate. The entropic C-function partially increases as a function of $L$ at finite temperature, while it is suppressed by large gluon condensate due to competitive behaviors of two.

\section{Summary and conclusions}\label{sec:05}
In this paper, we investigate the relationship among gluon condensate, holographic entanglement entropy, and phase transition. An Einstein-Dilaton model is utilized, and the dilaton field is connected to the gluon condensate. We first consider the case of vanishing temperature. It is found that the difference in entanglement entropy changes sign, and a phase transition from a connected to a disconnected surface occurs, corresponding to the confinement/deconfinement phase transition.

For large gluon condensate, the critical length shifts to the left, indicating that the system tends to be confined. We also analyze the finite part of the holographic entanglement entropy: \( S_A = S_{A,\text{div}} + S_{A,\text{fin}} \), where \( S_{A,\text{div}} \) depends on the cut-off scale, and \( S_{A,\text{fin}} \) does not. The gluon condensate slightly decreases the finite part of the connected surface. This suggests that the gluon condensate induces confinement and reduces the degrees of freedom of entangled states because quarks cannot be isolated. Indeed, the entropic C-function decreases as a function of the length \( L \) and drops to zero at the critical length where no entangled states remain.

Even at finite temperature, the holographic entanglement entropy exhibits a phase transition between connected and disconnected surfaces, which differs from the findings in \cite{Dudal:2018ztm,Mahapatra:2019uql}. When computing disconnected surfaces, we do not have the entropy contribution from the black hole horizon, which vanishes in the dilaton black hole solution. At finite temperature, the effect of the gluon condensate on the difference in entanglement entropy is qualitatively similar to the case of vanishing temperature, whereas temperature affects entanglement entropy in the opposite manner. The entropic C-function is useful for capturing physics at finite temperature. Moreover, the entropic C-function increases as a function of length \( L \) at finite temperature and captures thermal excitations. Thermal excitations exhibit competitive behaviors with the gluon condensate, as shown in Fig. \ref{TCS2}. Thus, we suggest that holographic entanglement entropy can serve as a useful probe for the confinement/deconfinement phase. Further studies in a realistic holographic model that captures more properties of QCD and the the case of finite chemical potential will be pursued in future research.

\section{Acknowledgments}
The authors thank Zi-qiang Zhang for his useful discussions. This work is supported by the National Science Foundation of China (NSFC) under grant numbers 12405154, 12175100. It is also supported by the Open Fund for Key Laboratories of the Ministry of Education under grant number QLPL2024P01.

\section*{References}

\bibliography{entanglement_entropy}

\begin{thebibliography}{91}%
\makeatletter
\providecommand \@ifxundefined [1]{%
 \@ifx{#1\undefined}
}%
\providecommand \@ifnum [1]{%
 \ifnum #1\expandafter \@firstoftwo
 \else \expandafter \@secondoftwo
 \fi
}%
\providecommand \@ifx [1]{%
 \ifx #1\expandafter \@firstoftwo
 \else \expandafter \@secondoftwo
 \fi
}%
\providecommand \natexlab [1]{#1}%
\providecommand \enquote  [1]{``#1''}%
\providecommand \bibnamefont  [1]{#1}%
\providecommand \bibfnamefont [1]{#1}%
\providecommand \citenamefont [1]{#1}%
\providecommand \href@noop [0]{\@secondoftwo}%
\providecommand \href [0]{\begingroup \@sanitize@url \@href}%
\providecommand \@href[1]{\@@startlink{#1}\@@href}%
\providecommand \@@href[1]{\endgroup#1\@@endlink}%
\providecommand \@sanitize@url [0]{\catcode `\\12\catcode `\$12\catcode
  `\&12\catcode `\#12\catcode `\^12\catcode `\_12\catcode `\%12\relax}%
\providecommand \@@startlink[1]{}%
\providecommand \@@endlink[0]{}%
\providecommand \url  [0]{\begingroup\@sanitize@url \@url }%
\providecommand \@url [1]{\endgroup\@href {#1}{\urlprefix }}%
\providecommand \urlprefix  [0]{URL }%
\providecommand \Eprint [0]{\href }%
\providecommand \doibase [0]{http://dx.doi.org/}%
\providecommand \selectlanguage [0]{\@gobble}%
\providecommand \bibinfo  [0]{\@secondoftwo}%
\providecommand \bibfield  [0]{\@secondoftwo}%
\providecommand \translation [1]{[#1]}%
\providecommand \BibitemOpen [0]{}%
\providecommand \bibitemStop [0]{}%
\providecommand \bibitemNoStop [0]{.\EOS\space}%
\providecommand \EOS [0]{\spacefactor3000\relax}%
\providecommand \BibitemShut  [1]{\csname bibitem#1\endcsname}%
\let\auto@bib@innerbib\@empty
\bibitem [{\citenamefont {Houston}\ and\ \citenamefont
  {Pottinger}(1979)}]{Houston:1979ia}%
  \BibitemOpen
  \bibfield  {author} {\bibinfo {author} {\bibfnamefont {P.}~\bibnamefont
  {Houston}}\ and\ \bibinfo {author} {\bibfnamefont {D.}~\bibnamefont
  {Pottinger}},\ }\href {\doibase 10.1007/BF01577402} {\bibfield  {journal}
  {\bibinfo  {journal} {Z. Phys. C}\ }\textbf {\bibinfo {volume} {3}},\
  \bibinfo {pages} {83} (\bibinfo {year} {1979})}\BibitemShut {NoStop}%
\bibitem [{\citenamefont {Wang}\ and\ \citenamefont {Li}(1989)}]{Wang:1989ag}%
  \BibitemOpen
  \bibfield  {author} {\bibinfo {author} {\bibfnamefont {E.-k.}\ \bibnamefont
  {Wang}}\ and\ \bibinfo {author} {\bibfnamefont {J.-r.}\ \bibnamefont {Li}},\
  }\href@noop {} {\bibfield  {journal} {\bibinfo  {journal} {HEPNP}\ }\textbf
  {\bibinfo {volume} {13}},\ \bibinfo {pages} {1003} (\bibinfo {year}
  {1989})}\BibitemShut {NoStop}%
\bibitem [{\citenamefont {Baldo}\ \emph {et~al.}(2004)\citenamefont {Baldo},
  \citenamefont {Castorina},\ and\ \citenamefont {Zappala}}]{Baldo:2003id}%
  \BibitemOpen
  \bibfield  {author} {\bibinfo {author} {\bibfnamefont {M.}~\bibnamefont
  {Baldo}}, \bibinfo {author} {\bibfnamefont {P.}~\bibnamefont {Castorina}}, \
  and\ \bibinfo {author} {\bibfnamefont {D.}~\bibnamefont {Zappala}},\ }\href
  {\doibase 10.1016/j.nuclphysa.2004.07.003} {\bibfield  {journal} {\bibinfo
  {journal} {Nucl. Phys. A}\ }\textbf {\bibinfo {volume} {743}},\ \bibinfo
  {pages} {3} (\bibinfo {year} {2004})},\ \Eprint
  {http://arxiv.org/abs/nucl-th/0311038} {arXiv:nucl-th/0311038} \BibitemShut
  {NoStop}%
\bibitem [{\citenamefont {Brown}\ \emph {et~al.}(2007)\citenamefont {Brown},
  \citenamefont {Holt}, \citenamefont {Lee},\ and\ \citenamefont
  {Rho}}]{Brown:2006vn}%
  \BibitemOpen
  \bibfield  {author} {\bibinfo {author} {\bibfnamefont {G.~E.}\ \bibnamefont
  {Brown}}, \bibinfo {author} {\bibfnamefont {J.~W.}\ \bibnamefont {Holt}},
  \bibinfo {author} {\bibfnamefont {C.-H.}\ \bibnamefont {Lee}}, \ and\
  \bibinfo {author} {\bibfnamefont {M.}~\bibnamefont {Rho}},\ }\href {\doibase
  10.1016/j.physrep.2006.12.002} {\bibfield  {journal} {\bibinfo  {journal}
  {Phys. Rept.}\ }\textbf {\bibinfo {volume} {439}},\ \bibinfo {pages} {161}
  (\bibinfo {year} {2007})},\ \Eprint {http://arxiv.org/abs/nucl-th/0608023}
  {arXiv:nucl-th/0608023} \BibitemShut {NoStop}%
\bibitem [{\citenamefont {Castorina}\ and\ \citenamefont
  {Mannarelli}(2007)}]{Castorina:2007qv}%
  \BibitemOpen
  \bibfield  {author} {\bibinfo {author} {\bibfnamefont {P.}~\bibnamefont
  {Castorina}}\ and\ \bibinfo {author} {\bibfnamefont {M.}~\bibnamefont
  {Mannarelli}},\ }\href {\doibase 10.1103/PhysRevC.75.054901} {\bibfield
  {journal} {\bibinfo  {journal} {Phys. Rev. C}\ }\textbf {\bibinfo {volume}
  {75}},\ \bibinfo {pages} {054901} (\bibinfo {year} {2007})},\ \Eprint
  {http://arxiv.org/abs/hep-ph/0701206} {arXiv:hep-ph/0701206} \BibitemShut
  {NoStop}%
\bibitem [{\citenamefont {Herzog}(2007)}]{Herzog:2006ra}%
  \BibitemOpen
  \bibfield  {author} {\bibinfo {author} {\bibfnamefont {C.~P.}\ \bibnamefont
  {Herzog}},\ }\href {\doibase 10.1103/PhysRevLett.98.091601} {\bibfield
  {journal} {\bibinfo  {journal} {Phys. Rev. Lett.}\ }\textbf {\bibinfo
  {volume} {98}},\ \bibinfo {pages} {091601} (\bibinfo {year} {2007})},\
  \Eprint {http://arxiv.org/abs/hep-th/0608151} {arXiv:hep-th/0608151}
  \BibitemShut {NoStop}%
\bibitem [{\citenamefont {Colangelo}\ \emph {et~al.}(2011)\citenamefont
  {Colangelo}, \citenamefont {Giannuzzi},\ and\ \citenamefont
  {Nicotri}}]{Colangelo:2010pe}%
  \BibitemOpen
  \bibfield  {author} {\bibinfo {author} {\bibfnamefont {P.}~\bibnamefont
  {Colangelo}}, \bibinfo {author} {\bibfnamefont {F.}~\bibnamefont
  {Giannuzzi}}, \ and\ \bibinfo {author} {\bibfnamefont {S.}~\bibnamefont
  {Nicotri}},\ }\href {\doibase 10.1103/PhysRevD.83.035015} {\bibfield
  {journal} {\bibinfo  {journal} {Phys. Rev. D}\ }\textbf {\bibinfo {volume}
  {83}},\ \bibinfo {pages} {035015} (\bibinfo {year} {2011})},\ \Eprint
  {http://arxiv.org/abs/1008.3116} {arXiv:1008.3116 [hep-ph]} \BibitemShut
  {NoStop}%
\bibitem [{\citenamefont {Kim}\ \emph {et~al.}(2009{\natexlab{a}})\citenamefont
  {Kim}, \citenamefont {Misumi},\ and\ \citenamefont {Shin}}]{Kim:2009wt}%
  \BibitemOpen
  \bibfield  {author} {\bibinfo {author} {\bibfnamefont {Y.}~\bibnamefont
  {Kim}}, \bibinfo {author} {\bibfnamefont {T.}~\bibnamefont {Misumi}}, \ and\
  \bibinfo {author} {\bibfnamefont {I.~J.}\ \bibnamefont {Shin}},\ }\href@noop
  {} {\  (\bibinfo {year} {2009}{\natexlab{a}})},\ \Eprint
  {http://arxiv.org/abs/0911.3205} {arXiv:0911.3205 [hep-ph]} \BibitemShut
  {NoStop}%
\bibitem [{\citenamefont {Evans}\ \emph
  {et~al.}(2012{\natexlab{a}})\citenamefont {Evans}, \citenamefont {Gebauer},
  \citenamefont {Magou},\ and\ \citenamefont {Kim}}]{Evans:2011eu}%
  \BibitemOpen
  \bibfield  {author} {\bibinfo {author} {\bibfnamefont {N.}~\bibnamefont
  {Evans}}, \bibinfo {author} {\bibfnamefont {A.}~\bibnamefont {Gebauer}},
  \bibinfo {author} {\bibfnamefont {M.}~\bibnamefont {Magou}}, \ and\ \bibinfo
  {author} {\bibfnamefont {K.-Y.}\ \bibnamefont {Kim}},\ }\href {\doibase
  10.1088/0954-3899/39/5/054005} {\bibfield  {journal} {\bibinfo  {journal} {J.
  Phys. G}\ }\textbf {\bibinfo {volume} {39}},\ \bibinfo {pages} {054005}
  (\bibinfo {year} {2012}{\natexlab{a}})},\ \Eprint
  {http://arxiv.org/abs/1109.2633} {arXiv:1109.2633 [hep-th]} \BibitemShut
  {NoStop}%
\bibitem [{\citenamefont {DeWolfe}\ \emph {et~al.}(2011)\citenamefont
  {DeWolfe}, \citenamefont {Gubser},\ and\ \citenamefont
  {Rosen}}]{DeWolfe:2010he}%
  \BibitemOpen
  \bibfield  {author} {\bibinfo {author} {\bibfnamefont {O.}~\bibnamefont
  {DeWolfe}}, \bibinfo {author} {\bibfnamefont {S.~S.}\ \bibnamefont {Gubser}},
  \ and\ \bibinfo {author} {\bibfnamefont {C.}~\bibnamefont {Rosen}},\ }\href
  {\doibase 10.1103/PhysRevD.83.086005} {\bibfield  {journal} {\bibinfo
  {journal} {Phys. Rev. D}\ }\textbf {\bibinfo {volume} {83}},\ \bibinfo
  {pages} {086005} (\bibinfo {year} {2011})},\ \Eprint
  {http://arxiv.org/abs/1012.1864} {arXiv:1012.1864 [hep-th]} \BibitemShut
  {NoStop}%
\bibitem [{\citenamefont {Li}\ \emph {et~al.}(2011)\citenamefont {Li},
  \citenamefont {He}, \citenamefont {Huang},\ and\ \citenamefont
  {Yan}}]{Li:2011hp}%
  \BibitemOpen
  \bibfield  {author} {\bibinfo {author} {\bibfnamefont {D.}~\bibnamefont
  {Li}}, \bibinfo {author} {\bibfnamefont {S.}~\bibnamefont {He}}, \bibinfo
  {author} {\bibfnamefont {M.}~\bibnamefont {Huang}}, \ and\ \bibinfo {author}
  {\bibfnamefont {Q.-S.}\ \bibnamefont {Yan}},\ }\href {\doibase
  10.1007/JHEP09(2011)041} {\bibfield  {journal} {\bibinfo  {journal} {JHEP}\
  }\textbf {\bibinfo {volume} {09}},\ \bibinfo {pages} {041} (\bibinfo {year}
  {2011})},\ \Eprint {http://arxiv.org/abs/1103.5389} {arXiv:1103.5389
  [hep-th]} \BibitemShut {NoStop}%
\bibitem [{\citenamefont {Evans}\ \emph
  {et~al.}(2012{\natexlab{b}})\citenamefont {Evans}, \citenamefont {Kim},
  \citenamefont {Magou}, \citenamefont {Seo},\ and\ \citenamefont
  {Sin}}]{Evans:2012cx}%
  \BibitemOpen
  \bibfield  {author} {\bibinfo {author} {\bibfnamefont {N.}~\bibnamefont
  {Evans}}, \bibinfo {author} {\bibfnamefont {K.-Y.}\ \bibnamefont {Kim}},
  \bibinfo {author} {\bibfnamefont {M.}~\bibnamefont {Magou}}, \bibinfo
  {author} {\bibfnamefont {Y.}~\bibnamefont {Seo}}, \ and\ \bibinfo {author}
  {\bibfnamefont {S.-J.}\ \bibnamefont {Sin}},\ }\href {\doibase
  10.1007/JHEP09(2012)045} {\bibfield  {journal} {\bibinfo  {journal} {JHEP}\
  }\textbf {\bibinfo {volume} {09}},\ \bibinfo {pages} {045} (\bibinfo {year}
  {2012}{\natexlab{b}})},\ \Eprint {http://arxiv.org/abs/1204.5640}
  {arXiv:1204.5640 [hep-th]} \BibitemShut {NoStop}%
\bibitem [{\citenamefont {Cai}\ \emph {et~al.}(2012)\citenamefont {Cai},
  \citenamefont {He},\ and\ \citenamefont {Li}}]{Cai:2012xh}%
  \BibitemOpen
  \bibfield  {author} {\bibinfo {author} {\bibfnamefont {R.-G.}\ \bibnamefont
  {Cai}}, \bibinfo {author} {\bibfnamefont {S.}~\bibnamefont {He}}, \ and\
  \bibinfo {author} {\bibfnamefont {D.}~\bibnamefont {Li}},\ }\href {\doibase
  10.1007/JHEP03(2012)033} {\bibfield  {journal} {\bibinfo  {journal} {JHEP}\
  }\textbf {\bibinfo {volume} {03}},\ \bibinfo {pages} {033} (\bibinfo {year}
  {2012})},\ \Eprint {http://arxiv.org/abs/1201.0820} {arXiv:1201.0820
  [hep-th]} \BibitemShut {NoStop}%
\bibitem [{\citenamefont {Cai}\ \emph {et~al.}(2013)\citenamefont {Cai},
  \citenamefont {Chakrabortty}, \citenamefont {He},\ and\ \citenamefont
  {Li}}]{Cai:2012eh}%
  \BibitemOpen
  \bibfield  {author} {\bibinfo {author} {\bibfnamefont {R.-G.}\ \bibnamefont
  {Cai}}, \bibinfo {author} {\bibfnamefont {S.}~\bibnamefont {Chakrabortty}},
  \bibinfo {author} {\bibfnamefont {S.}~\bibnamefont {He}}, \ and\ \bibinfo
  {author} {\bibfnamefont {L.}~\bibnamefont {Li}},\ }\href {\doibase
  10.1007/JHEP02(2013)068} {\bibfield  {journal} {\bibinfo  {journal} {JHEP}\
  }\textbf {\bibinfo {volume} {02}},\ \bibinfo {pages} {068} (\bibinfo {year}
  {2013})},\ \Eprint {http://arxiv.org/abs/1209.4512} {arXiv:1209.4512
  [hep-th]} \BibitemShut {NoStop}%
\bibitem [{\citenamefont {Alho}\ \emph {et~al.}(2013)\citenamefont {Alho},
  \citenamefont {J\"arvinen}, \citenamefont {Kajantie}, \citenamefont
  {Kiritsis},\ and\ \citenamefont {Tuominen}}]{Alho:2012mh}%
  \BibitemOpen
  \bibfield  {author} {\bibinfo {author} {\bibfnamefont {T.}~\bibnamefont
  {Alho}}, \bibinfo {author} {\bibfnamefont {M.}~\bibnamefont {J\"arvinen}},
  \bibinfo {author} {\bibfnamefont {K.}~\bibnamefont {Kajantie}}, \bibinfo
  {author} {\bibfnamefont {E.}~\bibnamefont {Kiritsis}}, \ and\ \bibinfo
  {author} {\bibfnamefont {K.}~\bibnamefont {Tuominen}},\ }\href {\doibase
  10.1007/JHEP01(2013)093} {\bibfield  {journal} {\bibinfo  {journal} {JHEP}\
  }\textbf {\bibinfo {volume} {01}},\ \bibinfo {pages} {093} (\bibinfo {year}
  {2013})},\ \Eprint {http://arxiv.org/abs/1210.4516} {arXiv:1210.4516
  [hep-ph]} \BibitemShut {NoStop}%
\bibitem [{\citenamefont {Yang}\ and\ \citenamefont
  {Yuan}(2015)}]{Yang:2015aia}%
  \BibitemOpen
  \bibfield  {author} {\bibinfo {author} {\bibfnamefont {Y.}~\bibnamefont
  {Yang}}\ and\ \bibinfo {author} {\bibfnamefont {P.-H.}\ \bibnamefont
  {Yuan}},\ }\href {\doibase 10.1007/JHEP12(2015)161} {\bibfield  {journal}
  {\bibinfo  {journal} {JHEP}\ }\textbf {\bibinfo {volume} {12}},\ \bibinfo
  {pages} {161} (\bibinfo {year} {2015})},\ \Eprint
  {http://arxiv.org/abs/1506.05930} {arXiv:1506.05930 [hep-th]} \BibitemShut
  {NoStop}%
\bibitem [{\citenamefont {Critelli}\ \emph {et~al.}(2016)\citenamefont
  {Critelli}, \citenamefont {Rougemont}, \citenamefont {Finazzo},\ and\
  \citenamefont {Noronha}}]{Critelli:2016cvq}%
  \BibitemOpen
  \bibfield  {author} {\bibinfo {author} {\bibfnamefont {R.}~\bibnamefont
  {Critelli}}, \bibinfo {author} {\bibfnamefont {R.}~\bibnamefont {Rougemont}},
  \bibinfo {author} {\bibfnamefont {S.~I.}\ \bibnamefont {Finazzo}}, \ and\
  \bibinfo {author} {\bibfnamefont {J.}~\bibnamefont {Noronha}},\ }\href
  {\doibase 10.1103/PhysRevD.94.125019} {\bibfield  {journal} {\bibinfo
  {journal} {Phys. Rev. D}\ }\textbf {\bibinfo {volume} {94}},\ \bibinfo
  {pages} {125019} (\bibinfo {year} {2016})},\ \Eprint
  {http://arxiv.org/abs/1606.09484} {arXiv:1606.09484 [hep-ph]} \BibitemShut
  {NoStop}%
\bibitem [{\citenamefont {Li}\ \emph {et~al.}(2017{\natexlab{a}})\citenamefont
  {Li}, \citenamefont {Yang},\ and\ \citenamefont {Yuan}}]{Li:2017tdz}%
  \BibitemOpen
  \bibfield  {author} {\bibinfo {author} {\bibfnamefont {M.-W.}\ \bibnamefont
  {Li}}, \bibinfo {author} {\bibfnamefont {Y.}~\bibnamefont {Yang}}, \ and\
  \bibinfo {author} {\bibfnamefont {P.-H.}\ \bibnamefont {Yuan}},\ }\href
  {\doibase 10.1103/PhysRevD.96.066013} {\bibfield  {journal} {\bibinfo
  {journal} {Phys. Rev. D}\ }\textbf {\bibinfo {volume} {96}},\ \bibinfo
  {pages} {066013} (\bibinfo {year} {2017}{\natexlab{a}})},\ \Eprint
  {http://arxiv.org/abs/1703.09184} {arXiv:1703.09184 [hep-th]} \BibitemShut
  {NoStop}%
\bibitem [{\citenamefont {Chelabi}\ \emph {et~al.}(2016)\citenamefont
  {Chelabi}, \citenamefont {Fang}, \citenamefont {Huang}, \citenamefont {Li},\
  and\ \citenamefont {Wu}}]{Chelabi:2015gpc}%
  \BibitemOpen
  \bibfield  {author} {\bibinfo {author} {\bibfnamefont {K.}~\bibnamefont
  {Chelabi}}, \bibinfo {author} {\bibfnamefont {Z.}~\bibnamefont {Fang}},
  \bibinfo {author} {\bibfnamefont {M.}~\bibnamefont {Huang}}, \bibinfo
  {author} {\bibfnamefont {D.}~\bibnamefont {Li}}, \ and\ \bibinfo {author}
  {\bibfnamefont {Y.-L.}\ \bibnamefont {Wu}},\ }\href {\doibase
  10.1007/JHEP04(2016)036} {\bibfield  {journal} {\bibinfo  {journal} {JHEP}\
  }\textbf {\bibinfo {volume} {04}},\ \bibinfo {pages} {036} (\bibinfo {year}
  {2016})},\ \Eprint {http://arxiv.org/abs/1512.06493} {arXiv:1512.06493
  [hep-ph]} \BibitemShut {NoStop}%
\bibitem [{\citenamefont {Li}\ \emph {et~al.}(2017{\natexlab{b}})\citenamefont
  {Li}, \citenamefont {Huang}, \citenamefont {Yang},\ and\ \citenamefont
  {Yuan}}]{Li:2016gfn}%
  \BibitemOpen
  \bibfield  {author} {\bibinfo {author} {\bibfnamefont {D.}~\bibnamefont
  {Li}}, \bibinfo {author} {\bibfnamefont {M.}~\bibnamefont {Huang}}, \bibinfo
  {author} {\bibfnamefont {Y.}~\bibnamefont {Yang}}, \ and\ \bibinfo {author}
  {\bibfnamefont {P.-H.}\ \bibnamefont {Yuan}},\ }\href {\doibase
  10.1007/JHEP02(2017)030} {\bibfield  {journal} {\bibinfo  {journal} {JHEP}\
  }\textbf {\bibinfo {volume} {02}},\ \bibinfo {pages} {030} (\bibinfo {year}
  {2017}{\natexlab{b}})},\ \Eprint {http://arxiv.org/abs/1610.04618}
  {arXiv:1610.04618 [hep-th]} \BibitemShut {NoStop}%
\bibitem [{\citenamefont {Li}\ and\ \citenamefont {Huang}(2017)}]{Li:2016smq}%
  \BibitemOpen
  \bibfield  {author} {\bibinfo {author} {\bibfnamefont {D.}~\bibnamefont
  {Li}}\ and\ \bibinfo {author} {\bibfnamefont {M.}~\bibnamefont {Huang}},\
  }\href {\doibase 10.1007/JHEP02(2017)042} {\bibfield  {journal} {\bibinfo
  {journal} {JHEP}\ }\textbf {\bibinfo {volume} {02}},\ \bibinfo {pages} {042}
  (\bibinfo {year} {2017})},\ \Eprint {http://arxiv.org/abs/1610.09814}
  {arXiv:1610.09814 [hep-ph]} \BibitemShut {NoStop}%
\bibitem [{\citenamefont {Li}\ \emph {et~al.}(2018)\citenamefont {Li},
  \citenamefont {Chen}, \citenamefont {Li},\ and\ \citenamefont
  {Huang}}]{Li:2017ple}%
  \BibitemOpen
  \bibfield  {author} {\bibinfo {author} {\bibfnamefont {Z.}~\bibnamefont
  {Li}}, \bibinfo {author} {\bibfnamefont {Y.}~\bibnamefont {Chen}}, \bibinfo
  {author} {\bibfnamefont {D.}~\bibnamefont {Li}}, \ and\ \bibinfo {author}
  {\bibfnamefont {M.}~\bibnamefont {Huang}},\ }\href {\doibase
  10.1088/1674-1137/42/1/013103} {\bibfield  {journal} {\bibinfo  {journal}
  {Chin. Phys. C}\ }\textbf {\bibinfo {volume} {42}},\ \bibinfo {pages}
  {013103} (\bibinfo {year} {2018})},\ \Eprint
  {http://arxiv.org/abs/1706.02238} {arXiv:1706.02238 [hep-ph]} \BibitemShut
  {NoStop}%
\bibitem [{\citenamefont {Knaute}\ \emph {et~al.}(2018)\citenamefont {Knaute},
  \citenamefont {Yaresko},\ and\ \citenamefont {K\"ampfer}}]{Knaute:2017opk}%
  \BibitemOpen
  \bibfield  {author} {\bibinfo {author} {\bibfnamefont {J.}~\bibnamefont
  {Knaute}}, \bibinfo {author} {\bibfnamefont {R.}~\bibnamefont {Yaresko}}, \
  and\ \bibinfo {author} {\bibfnamefont {B.}~\bibnamefont {K\"ampfer}},\ }\href
  {\doibase 10.1016/j.physletb.2018.01.053} {\bibfield  {journal} {\bibinfo
  {journal} {Phys. Lett. B}\ }\textbf {\bibinfo {volume} {778}},\ \bibinfo
  {pages} {419} (\bibinfo {year} {2018})},\ \Eprint
  {http://arxiv.org/abs/1702.06731} {arXiv:1702.06731 [hep-ph]} \BibitemShut
  {NoStop}%
\bibitem [{\citenamefont {Gursoy}\ \emph {et~al.}(2018)\citenamefont {Gursoy},
  \citenamefont {Jarvinen},\ and\ \citenamefont {Nijs}}]{Gursoy:2017wzz}%
  \BibitemOpen
  \bibfield  {author} {\bibinfo {author} {\bibfnamefont {U.}~\bibnamefont
  {Gursoy}}, \bibinfo {author} {\bibfnamefont {M.}~\bibnamefont {Jarvinen}}, \
  and\ \bibinfo {author} {\bibfnamefont {G.}~\bibnamefont {Nijs}},\ }\href
  {\doibase 10.1103/PhysRevLett.120.242002} {\bibfield  {journal} {\bibinfo
  {journal} {Phys. Rev. Lett.}\ }\textbf {\bibinfo {volume} {120}},\ \bibinfo
  {pages} {242002} (\bibinfo {year} {2018})},\ \Eprint
  {http://arxiv.org/abs/1707.00872} {arXiv:1707.00872 [hep-th]} \BibitemShut
  {NoStop}%
\bibitem [{\citenamefont {G\"ursoy}\ \emph {et~al.}(2019)\citenamefont
  {G\"ursoy}, \citenamefont {J\"arvinen}, \citenamefont {Nijs},\ and\
  \citenamefont {Pedraza}}]{Gursoy:2018ydr}%
  \BibitemOpen
  \bibfield  {author} {\bibinfo {author} {\bibfnamefont {U.}~\bibnamefont
  {G\"ursoy}}, \bibinfo {author} {\bibfnamefont {M.}~\bibnamefont
  {J\"arvinen}}, \bibinfo {author} {\bibfnamefont {G.}~\bibnamefont {Nijs}}, \
  and\ \bibinfo {author} {\bibfnamefont {J.~F.}\ \bibnamefont {Pedraza}},\
  }\href {\doibase 10.1007/JHEP04(2019)071} {\bibfield  {journal} {\bibinfo
  {journal} {JHEP}\ }\textbf {\bibinfo {volume} {04}},\ \bibinfo {pages} {071}
  (\bibinfo {year} {2019})},\ \bibinfo {note} {[Erratum: JHEP 09, 059
  (2020)]},\ \Eprint {http://arxiv.org/abs/1811.11724} {arXiv:1811.11724
  [hep-th]} \BibitemShut {NoStop}%
\bibitem [{\citenamefont {Chen}\ \emph {et~al.}(2019)\citenamefont {Chen},
  \citenamefont {Li},\ and\ \citenamefont {Huang}}]{Chen:2018vty}%
  \BibitemOpen
  \bibfield  {author} {\bibinfo {author} {\bibfnamefont {X.}~\bibnamefont
  {Chen}}, \bibinfo {author} {\bibfnamefont {D.}~\bibnamefont {Li}}, \ and\
  \bibinfo {author} {\bibfnamefont {M.}~\bibnamefont {Huang}},\ }\href
  {\doibase 10.1088/1674-1137/43/2/023105} {\bibfield  {journal} {\bibinfo
  {journal} {Chin. Phys. C}\ }\textbf {\bibinfo {volume} {43}},\ \bibinfo
  {pages} {023105} (\bibinfo {year} {2019})},\ \Eprint
  {http://arxiv.org/abs/1810.02136} {arXiv:1810.02136 [hep-ph]} \BibitemShut
  {NoStop}%
\bibitem [{\citenamefont {Chen}\ \emph {et~al.}(2020)\citenamefont {Chen},
  \citenamefont {Li}, \citenamefont {Hou},\ and\ \citenamefont
  {Huang}}]{Chen:2019rez}%
  \BibitemOpen
  \bibfield  {author} {\bibinfo {author} {\bibfnamefont {X.}~\bibnamefont
  {Chen}}, \bibinfo {author} {\bibfnamefont {D.}~\bibnamefont {Li}}, \bibinfo
  {author} {\bibfnamefont {D.}~\bibnamefont {Hou}}, \ and\ \bibinfo {author}
  {\bibfnamefont {M.}~\bibnamefont {Huang}},\ }\href {\doibase
  10.1007/JHEP03(2020)073} {\bibfield  {journal} {\bibinfo  {journal} {JHEP}\
  }\textbf {\bibinfo {volume} {03}},\ \bibinfo {pages} {073} (\bibinfo {year}
  {2020})},\ \Eprint {http://arxiv.org/abs/1908.02000} {arXiv:1908.02000
  [hep-ph]} \BibitemShut {NoStop}%
\bibitem [{\citenamefont {Chen}\ \emph {et~al.}(2018)\citenamefont {Chen},
  \citenamefont {Feng}, \citenamefont {Shi},\ and\ \citenamefont
  {Zhong}}]{Chen:2017lsf}%
  \BibitemOpen
  \bibfield  {author} {\bibinfo {author} {\bibfnamefont {X.}~\bibnamefont
  {Chen}}, \bibinfo {author} {\bibfnamefont {S.-Q.}\ \bibnamefont {Feng}},
  \bibinfo {author} {\bibfnamefont {Y.-F.}\ \bibnamefont {Shi}}, \ and\
  \bibinfo {author} {\bibfnamefont {Y.}~\bibnamefont {Zhong}},\ }\href
  {\doibase 10.1103/PhysRevD.97.066015} {\bibfield  {journal} {\bibinfo
  {journal} {Phys. Rev. D}\ }\textbf {\bibinfo {volume} {97}},\ \bibinfo
  {pages} {066015} (\bibinfo {year} {2018})},\ \Eprint
  {http://arxiv.org/abs/1710.00465} {arXiv:1710.00465 [hep-ph]} \BibitemShut
  {NoStop}%
\bibitem [{\citenamefont {Cao}\ \emph {et~al.}(2020)\citenamefont {Cao},
  \citenamefont {Liu}, \citenamefont {Li},\ and\ \citenamefont
  {Ou}}]{Cao:2020ske}%
  \BibitemOpen
  \bibfield  {author} {\bibinfo {author} {\bibfnamefont {X.}~\bibnamefont
  {Cao}}, \bibinfo {author} {\bibfnamefont {H.}~\bibnamefont {Liu}}, \bibinfo
  {author} {\bibfnamefont {D.}~\bibnamefont {Li}}, \ and\ \bibinfo {author}
  {\bibfnamefont {G.}~\bibnamefont {Ou}},\ }\href {\doibase
  10.1088/1674-1137/44/8/083106} {\bibfield  {journal} {\bibinfo  {journal}
  {Chin. Phys. C}\ }\textbf {\bibinfo {volume} {44}},\ \bibinfo {pages}
  {083106} (\bibinfo {year} {2020})},\ \Eprint
  {http://arxiv.org/abs/2001.02888} {arXiv:2001.02888 [hep-ph]} \BibitemShut
  {NoStop}%
\bibitem [{\citenamefont {Rodrigues}\ \emph {et~al.}(2021)\citenamefont
  {Rodrigues}, \citenamefont {Li}, \citenamefont {Folco~Capossoli},\ and\
  \citenamefont {Boschi-Filho}}]{Rodrigues:2020ndy}%
  \BibitemOpen
  \bibfield  {author} {\bibinfo {author} {\bibfnamefont {D.~M.}\ \bibnamefont
  {Rodrigues}}, \bibinfo {author} {\bibfnamefont {D.}~\bibnamefont {Li}},
  \bibinfo {author} {\bibfnamefont {E.}~\bibnamefont {Folco~Capossoli}}, \ and\
  \bibinfo {author} {\bibfnamefont {H.}~\bibnamefont {Boschi-Filho}},\ }\href
  {\doibase 10.1103/PhysRevD.103.066022} {\bibfield  {journal} {\bibinfo
  {journal} {Phys. Rev. D}\ }\textbf {\bibinfo {volume} {103}},\ \bibinfo
  {pages} {066022} (\bibinfo {year} {2021})},\ \Eprint
  {http://arxiv.org/abs/2010.06762} {arXiv:2010.06762 [hep-th]} \BibitemShut
  {NoStop}%
\bibitem [{\citenamefont {Aref'eva}\ \emph {et~al.}(2021)\citenamefont
  {Aref'eva}, \citenamefont {Rannu},\ and\ \citenamefont
  {Slepov}}]{Arefeva:2020vae}%
  \BibitemOpen
  \bibfield  {author} {\bibinfo {author} {\bibfnamefont {I.~Y.}\ \bibnamefont
  {Aref'eva}}, \bibinfo {author} {\bibfnamefont {K.}~\bibnamefont {Rannu}}, \
  and\ \bibinfo {author} {\bibfnamefont {P.}~\bibnamefont {Slepov}},\ }\href
  {\doibase 10.1007/JHEP07(2021)161} {\bibfield  {journal} {\bibinfo  {journal}
  {JHEP}\ }\textbf {\bibinfo {volume} {07}},\ \bibinfo {pages} {161} (\bibinfo
  {year} {2021})},\ \Eprint {http://arxiv.org/abs/2011.07023} {arXiv:2011.07023
  [hep-th]} \BibitemShut {NoStop}%
\bibitem [{\citenamefont {Chen}\ \emph {et~al.}(2021)\citenamefont {Chen},
  \citenamefont {Zhang}, \citenamefont {Li}, \citenamefont {Hou},\ and\
  \citenamefont {Huang}}]{Chen:2020ath}%
  \BibitemOpen
  \bibfield  {author} {\bibinfo {author} {\bibfnamefont {X.}~\bibnamefont
  {Chen}}, \bibinfo {author} {\bibfnamefont {L.}~\bibnamefont {Zhang}},
  \bibinfo {author} {\bibfnamefont {D.}~\bibnamefont {Li}}, \bibinfo {author}
  {\bibfnamefont {D.}~\bibnamefont {Hou}}, \ and\ \bibinfo {author}
  {\bibfnamefont {M.}~\bibnamefont {Huang}},\ }\href {\doibase
  10.1007/JHEP07(2021)132} {\bibfield  {journal} {\bibinfo  {journal} {JHEP}\
  }\textbf {\bibinfo {volume} {07}},\ \bibinfo {pages} {132} (\bibinfo {year}
  {2021})},\ \Eprint {http://arxiv.org/abs/2010.14478} {arXiv:2010.14478
  [hep-ph]} \BibitemShut {NoStop}%
\bibitem [{\citenamefont {Zhu}\ \emph {et~al.}(2019)\citenamefont {Zhu},
  \citenamefont {Feng}, \citenamefont {Shi},\ and\ \citenamefont
  {Zhong}}]{Zhu:2019ujc}%
  \BibitemOpen
  \bibfield  {author} {\bibinfo {author} {\bibfnamefont {Z.-R.}\ \bibnamefont
  {Zhu}}, \bibinfo {author} {\bibfnamefont {S.-Q.}\ \bibnamefont {Feng}},
  \bibinfo {author} {\bibfnamefont {Y.-F.}\ \bibnamefont {Shi}}, \ and\
  \bibinfo {author} {\bibfnamefont {Y.}~\bibnamefont {Zhong}},\ }\href
  {\doibase 10.1103/PhysRevD.99.126001} {\bibfield  {journal} {\bibinfo
  {journal} {Phys. Rev. D}\ }\textbf {\bibinfo {volume} {99}},\ \bibinfo
  {pages} {126001} (\bibinfo {year} {2019})},\ \Eprint
  {http://arxiv.org/abs/1901.09304} {arXiv:1901.09304 [hep-ph]} \BibitemShut
  {NoStop}%
\bibitem [{\citenamefont {Zhu}\ \emph {et~al.}(2020)\citenamefont {Zhu},
  \citenamefont {Hou},\ and\ \citenamefont {Chen}}]{Zhu:2019igg}%
  \BibitemOpen
  \bibfield  {author} {\bibinfo {author} {\bibfnamefont {Z.-R.}\ \bibnamefont
  {Zhu}}, \bibinfo {author} {\bibfnamefont {D.-f.}\ \bibnamefont {Hou}}, \ and\
  \bibinfo {author} {\bibfnamefont {X.}~\bibnamefont {Chen}},\ }\href {\doibase
  10.1140/epjc/s10052-020-8110-8} {\bibfield  {journal} {\bibinfo  {journal}
  {Eur. Phys. J. C}\ }\textbf {\bibinfo {volume} {80}},\ \bibinfo {pages} {550}
  (\bibinfo {year} {2020})},\ \Eprint {http://arxiv.org/abs/1912.05806}
  {arXiv:1912.05806 [hep-ph]} \BibitemShut {NoStop}%
\bibitem [{\citenamefont {Zhu}\ \emph {et~al.}(2021)\citenamefont {Zhu},
  \citenamefont {Liu},\ and\ \citenamefont {Hou}}]{Zhu:2021ucv}%
  \BibitemOpen
  \bibfield  {author} {\bibinfo {author} {\bibfnamefont {Z.-R.}\ \bibnamefont
  {Zhu}}, \bibinfo {author} {\bibfnamefont {Y.-K.}\ \bibnamefont {Liu}}, \ and\
  \bibinfo {author} {\bibfnamefont {D.}~\bibnamefont {Hou}},\ }\href@noop {} {\
   (\bibinfo {year} {2021})},\ \Eprint {http://arxiv.org/abs/2108.05148}
  {arXiv:2108.05148 [hep-ph]} \BibitemShut {NoStop}%
\bibitem [{\citenamefont {Giataganas}\ \emph {et~al.}(2018)\citenamefont
  {Giataganas}, \citenamefont {G\"ursoy},\ and\ \citenamefont
  {Pedraza}}]{Giataganas:2017koz}%
  \BibitemOpen
  \bibfield  {author} {\bibinfo {author} {\bibfnamefont {D.}~\bibnamefont
  {Giataganas}}, \bibinfo {author} {\bibfnamefont {U.}~\bibnamefont
  {G\"ursoy}}, \ and\ \bibinfo {author} {\bibfnamefont {J.~F.}\ \bibnamefont
  {Pedraza}},\ }\href {\doibase 10.1103/PhysRevLett.121.121601} {\bibfield
  {journal} {\bibinfo  {journal} {Phys. Rev. Lett.}\ }\textbf {\bibinfo
  {volume} {121}},\ \bibinfo {pages} {121601} (\bibinfo {year} {2018})},\
  \Eprint {http://arxiv.org/abs/1708.05691} {arXiv:1708.05691 [hep-th]}
  \BibitemShut {NoStop}%
\bibitem [{\citenamefont {Nakas}\ and\ \citenamefont
  {Rigatos}(2020)}]{Nakas:2020hyo}%
  \BibitemOpen
  \bibfield  {author} {\bibinfo {author} {\bibfnamefont {T.}~\bibnamefont
  {Nakas}}\ and\ \bibinfo {author} {\bibfnamefont {K.~S.}\ \bibnamefont
  {Rigatos}},\ }\href {\doibase 10.1007/JHEP12(2020)157} {\bibfield  {journal}
  {\bibinfo  {journal} {JHEP}\ }\textbf {\bibinfo {volume} {12}},\ \bibinfo
  {pages} {157} (\bibinfo {year} {2020})},\ \Eprint
  {http://arxiv.org/abs/2010.00025} {arXiv:2010.00025 [hep-th]} \BibitemShut
  {NoStop}%
\bibitem [{\citenamefont {Abt}\ \emph {et~al.}(2019)\citenamefont {Abt},
  \citenamefont {Erdmenger}, \citenamefont {Evans},\ and\ \citenamefont
  {Rigatos}}]{Abt:2019tas}%
  \BibitemOpen
  \bibfield  {author} {\bibinfo {author} {\bibfnamefont {R.}~\bibnamefont
  {Abt}}, \bibinfo {author} {\bibfnamefont {J.}~\bibnamefont {Erdmenger}},
  \bibinfo {author} {\bibfnamefont {N.}~\bibnamefont {Evans}}, \ and\ \bibinfo
  {author} {\bibfnamefont {K.~S.}\ \bibnamefont {Rigatos}},\ }\href {\doibase
  10.1007/JHEP11(2019)160} {\bibfield  {journal} {\bibinfo  {journal} {JHEP}\
  }\textbf {\bibinfo {volume} {11}},\ \bibinfo {pages} {160} (\bibinfo {year}
  {2019})},\ \Eprint {http://arxiv.org/abs/1907.09489} {arXiv:1907.09489
  [hep-th]} \BibitemShut {NoStop}%
\bibitem [{\citenamefont {Csaki}\ and\ \citenamefont
  {Reece}(2007)}]{Csaki:2006ji}%
  \BibitemOpen
  \bibfield  {author} {\bibinfo {author} {\bibfnamefont {C.}~\bibnamefont
  {Csaki}}\ and\ \bibinfo {author} {\bibfnamefont {M.}~\bibnamefont {Reece}},\
  }\href {\doibase 10.1088/1126-6708/2007/05/062} {\bibfield  {journal}
  {\bibinfo  {journal} {JHEP}\ }\textbf {\bibinfo {volume} {05}},\ \bibinfo
  {pages} {062} (\bibinfo {year} {2007})},\ \Eprint
  {http://arxiv.org/abs/hep-ph/0608266} {arXiv:hep-ph/0608266} \BibitemShut
  {NoStop}%
\bibitem [{\citenamefont {Gubser}\ and\ \citenamefont
  {Nellore}(2008)}]{Gubser:2008ny}%
  \BibitemOpen
  \bibfield  {author} {\bibinfo {author} {\bibfnamefont {S.~S.}\ \bibnamefont
  {Gubser}}\ and\ \bibinfo {author} {\bibfnamefont {A.}~\bibnamefont
  {Nellore}},\ }\href {\doibase 10.1103/PhysRevD.78.086007} {\bibfield
  {journal} {\bibinfo  {journal} {Phys. Rev. D}\ }\textbf {\bibinfo {volume}
  {78}},\ \bibinfo {pages} {086007} (\bibinfo {year} {2008})},\ \Eprint
  {http://arxiv.org/abs/0804.0434} {arXiv:0804.0434 [hep-th]} \BibitemShut
  {NoStop}%
\bibitem [{\citenamefont {Shifman}\ \emph {et~al.}(1979)\citenamefont
  {Shifman}, \citenamefont {Vainshtein},\ and\ \citenamefont
  {Zakharov}}]{Shifman:1978bx}%
  \BibitemOpen
  \bibfield  {author} {\bibinfo {author} {\bibfnamefont {M.~A.}\ \bibnamefont
  {Shifman}}, \bibinfo {author} {\bibfnamefont {A.~I.}\ \bibnamefont
  {Vainshtein}}, \ and\ \bibinfo {author} {\bibfnamefont {V.~I.}\ \bibnamefont
  {Zakharov}},\ }\href {\doibase 10.1016/0550-3213(79)90022-1} {\bibfield
  {journal} {\bibinfo  {journal} {Nucl. Phys. B}\ }\textbf {\bibinfo {volume}
  {147}},\ \bibinfo {pages} {385} (\bibinfo {year} {1979})}\BibitemShut
  {NoStop}%
\bibitem [{\citenamefont {Lee}(1989)}]{Lee:1989qj}%
  \BibitemOpen
  \bibfield  {author} {\bibinfo {author} {\bibfnamefont {S.~H.}\ \bibnamefont
  {Lee}},\ }\href {\doibase 10.1103/PhysRevD.40.2484} {\bibfield  {journal}
  {\bibinfo  {journal} {Phys. Rev. D}\ }\textbf {\bibinfo {volume} {40}},\
  \bibinfo {pages} {2484} (\bibinfo {year} {1989})}\BibitemShut {NoStop}%
\bibitem [{\citenamefont {D'Elia}\ \emph {et~al.}(2003)\citenamefont {D'Elia},
  \citenamefont {Di~Giacomo},\ and\ \citenamefont
  {Meggiolaro}}]{DElia:2002hkf}%
  \BibitemOpen
  \bibfield  {author} {\bibinfo {author} {\bibfnamefont {M.}~\bibnamefont
  {D'Elia}}, \bibinfo {author} {\bibfnamefont {A.}~\bibnamefont {Di~Giacomo}},
  \ and\ \bibinfo {author} {\bibfnamefont {E.}~\bibnamefont {Meggiolaro}},\
  }\href {\doibase 10.1103/PhysRevD.67.114504} {\bibfield  {journal} {\bibinfo
  {journal} {Phys. Rev. D}\ }\textbf {\bibinfo {volume} {67}},\ \bibinfo
  {pages} {114504} (\bibinfo {year} {2003})},\ \Eprint
  {http://arxiv.org/abs/hep-lat/0205018} {arXiv:hep-lat/0205018} \BibitemShut
  {NoStop}%
\bibitem [{\citenamefont {Miller}(2007)}]{Miller:2006hr}%
  \BibitemOpen
  \bibfield  {author} {\bibinfo {author} {\bibfnamefont {D.~E.}\ \bibnamefont
  {Miller}},\ }\href {\doibase 10.1016/j.physrep.2007.02.012} {\bibfield
  {journal} {\bibinfo  {journal} {Phys. Rept.}\ }\textbf {\bibinfo {volume}
  {443}},\ \bibinfo {pages} {55} (\bibinfo {year} {2007})},\ \Eprint
  {http://arxiv.org/abs/hep-ph/0608234} {arXiv:hep-ph/0608234} \BibitemShut
  {NoStop}%
\bibitem [{\citenamefont {Nojiri}\ and\ \citenamefont
  {Odintsov}(1999)}]{Nojiri:1998yx}%
  \BibitemOpen
  \bibfield  {author} {\bibinfo {author} {\bibfnamefont {S.}~\bibnamefont
  {Nojiri}}\ and\ \bibinfo {author} {\bibfnamefont {S.~D.}\ \bibnamefont
  {Odintsov}},\ }\href {\doibase 10.1016/S0370-2693(99)00048-9} {\bibfield
  {journal} {\bibinfo  {journal} {Phys. Lett. B}\ }\textbf {\bibinfo {volume}
  {449}},\ \bibinfo {pages} {39} (\bibinfo {year} {1999})},\ \Eprint
  {http://arxiv.org/abs/hep-th/9812017} {arXiv:hep-th/9812017} \BibitemShut
  {NoStop}%
\bibitem [{\citenamefont {Gubser}(1999)}]{Gubser:1999pk}%
  \BibitemOpen
  \bibfield  {author} {\bibinfo {author} {\bibfnamefont {S.~S.}\ \bibnamefont
  {Gubser}},\ }\href@noop {} {\  (\bibinfo {year} {1999})},\ \Eprint
  {http://arxiv.org/abs/hep-th/9902155} {arXiv:hep-th/9902155} \BibitemShut
  {NoStop}%
\bibitem [{\citenamefont {Kehagias}\ and\ \citenamefont
  {Sfetsos}(1999)}]{Kehagias:1999tr}%
  \BibitemOpen
  \bibfield  {author} {\bibinfo {author} {\bibfnamefont {A.}~\bibnamefont
  {Kehagias}}\ and\ \bibinfo {author} {\bibfnamefont {K.}~\bibnamefont
  {Sfetsos}},\ }\href {\doibase 10.1016/S0370-2693(99)00393-7} {\bibfield
  {journal} {\bibinfo  {journal} {Phys. Lett. B}\ }\textbf {\bibinfo {volume}
  {454}},\ \bibinfo {pages} {270} (\bibinfo {year} {1999})},\ \Eprint
  {http://arxiv.org/abs/hep-th/9902125} {arXiv:hep-th/9902125} \BibitemShut
  {NoStop}%
\bibitem [{\citenamefont {Ko}\ \emph {et~al.}(2010)\citenamefont {Ko},
  \citenamefont {Lee},\ and\ \citenamefont {Park}}]{Ko:2009jc}%
  \BibitemOpen
  \bibfield  {author} {\bibinfo {author} {\bibfnamefont {Y.}~\bibnamefont
  {Ko}}, \bibinfo {author} {\bibfnamefont {B.-H.}\ \bibnamefont {Lee}}, \ and\
  \bibinfo {author} {\bibfnamefont {C.}~\bibnamefont {Park}},\ }\href {\doibase
  10.1007/JHEP04(2010)037} {\bibfield  {journal} {\bibinfo  {journal} {JHEP}\
  }\textbf {\bibinfo {volume} {04}},\ \bibinfo {pages} {037} (\bibinfo {year}
  {2010})},\ \Eprint {http://arxiv.org/abs/0912.5274} {arXiv:0912.5274
  [hep-ph]} \BibitemShut {NoStop}%
\bibitem [{\citenamefont {Kim}\ \emph {et~al.}(2009{\natexlab{b}})\citenamefont
  {Kim}, \citenamefont {Lee}, \citenamefont {Park},\ and\ \citenamefont
  {Sin}}]{Kim:2008ax}%
  \BibitemOpen
  \bibfield  {author} {\bibinfo {author} {\bibfnamefont {Y.}~\bibnamefont
  {Kim}}, \bibinfo {author} {\bibfnamefont {B.-H.}\ \bibnamefont {Lee}},
  \bibinfo {author} {\bibfnamefont {C.}~\bibnamefont {Park}}, \ and\ \bibinfo
  {author} {\bibfnamefont {S.-J.}\ \bibnamefont {Sin}},\ }\href {\doibase
  10.1103/PhysRevD.80.105016} {\bibfield  {journal} {\bibinfo  {journal} {Phys.
  Rev. D}\ }\textbf {\bibinfo {volume} {80}},\ \bibinfo {pages} {105016}
  (\bibinfo {year} {2009}{\natexlab{b}})},\ \Eprint
  {http://arxiv.org/abs/0808.1143} {arXiv:0808.1143 [hep-th]} \BibitemShut
  {NoStop}%
\bibitem [{\citenamefont {Zhao}\ \emph {et~al.}(2020)\citenamefont {Zhao},
  \citenamefont {Zhu},\ and\ \citenamefont {Chen}}]{Zhao:2019tjq}%
  \BibitemOpen
  \bibfield  {author} {\bibinfo {author} {\bibfnamefont {Y.-Q.}\ \bibnamefont
  {Zhao}}, \bibinfo {author} {\bibfnamefont {Z.-R.}\ \bibnamefont {Zhu}}, \
  and\ \bibinfo {author} {\bibfnamefont {X.}~\bibnamefont {Chen}},\ }\href
  {\doibase 10.1140/epja/s10050-020-00072-5} {\bibfield  {journal} {\bibinfo
  {journal} {Eur. Phys. J. A}\ }\textbf {\bibinfo {volume} {56}},\ \bibinfo
  {pages} {57} (\bibinfo {year} {2020})},\ \Eprint
  {http://arxiv.org/abs/1909.04994} {arXiv:1909.04994 [hep-ph]} \BibitemShut
  {NoStop}%
\bibitem [{\citenamefont {Zhang}\ and\ \citenamefont
  {Hou}(2020)}]{Zhang:2020upv}%
  \BibitemOpen
  \bibfield  {author} {\bibinfo {author} {\bibfnamefont {Z.-q.}\ \bibnamefont
  {Zhang}}\ and\ \bibinfo {author} {\bibfnamefont {D.-f.}\ \bibnamefont
  {Hou}},\ }\href {\doibase 10.1016/j.physletb.2020.135301} {\bibfield
  {journal} {\bibinfo  {journal} {Phys. Lett. B}\ }\textbf {\bibinfo {volume}
  {803}},\ \bibinfo {pages} {135301} (\bibinfo {year} {2020})}\BibitemShut
  {NoStop}%
\bibitem [{\citenamefont {Zhang}\ \emph {et~al.}(2020)\citenamefont {Zhang},
  \citenamefont {Zhu},\ and\ \citenamefont {Hou}}]{Zhang:2020noe}%
  \BibitemOpen
  \bibfield  {author} {\bibinfo {author} {\bibfnamefont {Z.-Q.}\ \bibnamefont
  {Zhang}}, \bibinfo {author} {\bibfnamefont {X.}~\bibnamefont {Zhu}}, \ and\
  \bibinfo {author} {\bibfnamefont {D.-F.}\ \bibnamefont {Hou}},\ }\href
  {\doibase 10.1103/PhysRevD.101.026017} {\bibfield  {journal} {\bibinfo
  {journal} {Phys. Rev. D}\ }\textbf {\bibinfo {volume} {101}},\ \bibinfo
  {pages} {026017} (\bibinfo {year} {2020})},\ \Eprint
  {http://arxiv.org/abs/2001.02321} {arXiv:2001.02321 [hep-th]} \BibitemShut
  {NoStop}%
\bibitem [{\citenamefont {Zhang}(2019)}]{Zhang:2019gki}%
  \BibitemOpen
  \bibfield  {author} {\bibinfo {author} {\bibfnamefont {Z.-q.}\ \bibnamefont
  {Zhang}},\ }\href {\doibase 10.1140/epjc/s10052-019-7503-z} {\bibfield
  {journal} {\bibinfo  {journal} {Eur. Phys. J. C}\ }\textbf {\bibinfo {volume}
  {79}},\ \bibinfo {pages} {992} (\bibinfo {year} {2019})}\BibitemShut
  {NoStop}%
\bibitem [{\citenamefont {Kim}\ \emph {et~al.}(2007)\citenamefont {Kim},
  \citenamefont {Lee}, \citenamefont {Park},\ and\ \citenamefont
  {Sin}}]{Kim:2007qk}%
  \BibitemOpen
  \bibfield  {author} {\bibinfo {author} {\bibfnamefont {Y.}~\bibnamefont
  {Kim}}, \bibinfo {author} {\bibfnamefont {B.-H.}\ \bibnamefont {Lee}},
  \bibinfo {author} {\bibfnamefont {C.}~\bibnamefont {Park}}, \ and\ \bibinfo
  {author} {\bibfnamefont {S.-J.}\ \bibnamefont {Sin}},\ }\href {\doibase
  10.1088/1126-6708/2007/09/105} {\bibfield  {journal} {\bibinfo  {journal}
  {JHEP}\ }\textbf {\bibinfo {volume} {09}},\ \bibinfo {pages} {105} (\bibinfo
  {year} {2007})},\ \Eprint {http://arxiv.org/abs/hep-th/0702131}
  {arXiv:hep-th/0702131} \BibitemShut {NoStop}%
\bibitem [{\citenamefont {Afonin}(2020)}]{Afonin:2020crk}%
  \BibitemOpen
  \bibfield  {author} {\bibinfo {author} {\bibfnamefont {S.~S.}\ \bibnamefont
  {Afonin}},\ }\href {\doibase 10.1016/j.physletb.2020.135780} {\bibfield
  {journal} {\bibinfo  {journal} {Phys. Lett. B}\ }\textbf {\bibinfo {volume}
  {809}},\ \bibinfo {pages} {135780} (\bibinfo {year} {2020})},\ \Eprint
  {http://arxiv.org/abs/2005.01550} {arXiv:2005.01550 [hep-ph]} \BibitemShut
  {NoStop}%
\bibitem [{\citenamefont {Ryu}\ and\ \citenamefont
  {Takayanagi}(2006{\natexlab{a}})}]{Ryu:2006bv}%
  \BibitemOpen
  \bibfield  {author} {\bibinfo {author} {\bibfnamefont {S.}~\bibnamefont
  {Ryu}}\ and\ \bibinfo {author} {\bibfnamefont {T.}~\bibnamefont
  {Takayanagi}},\ }\href {\doibase 10.1103/PhysRevLett.96.181602} {\bibfield
  {journal} {\bibinfo  {journal} {Phys. Rev. Lett.}\ }\textbf {\bibinfo
  {volume} {96}},\ \bibinfo {pages} {181602} (\bibinfo {year}
  {2006}{\natexlab{a}})},\ \Eprint {http://arxiv.org/abs/hep-th/0603001}
  {arXiv:hep-th/0603001} \BibitemShut {NoStop}%
\bibitem [{\citenamefont {Ryu}\ and\ \citenamefont
  {Takayanagi}(2006{\natexlab{b}})}]{Ryu:2006ef}%
  \BibitemOpen
  \bibfield  {author} {\bibinfo {author} {\bibfnamefont {S.}~\bibnamefont
  {Ryu}}\ and\ \bibinfo {author} {\bibfnamefont {T.}~\bibnamefont
  {Takayanagi}},\ }\href {\doibase 10.1088/1126-6708/2006/08/045} {\bibfield
  {journal} {\bibinfo  {journal} {JHEP}\ }\textbf {\bibinfo {volume} {08}},\
  \bibinfo {pages} {045} (\bibinfo {year} {2006}{\natexlab{b}})},\ \Eprint
  {http://arxiv.org/abs/hep-th/0605073} {arXiv:hep-th/0605073} \BibitemShut
  {NoStop}%
\bibitem [{\citenamefont {Nishioka}\ \emph {et~al.}(2009)\citenamefont
  {Nishioka}, \citenamefont {Ryu},\ and\ \citenamefont
  {Takayanagi}}]{Nishioka:2009un}%
  \BibitemOpen
  \bibfield  {author} {\bibinfo {author} {\bibfnamefont {T.}~\bibnamefont
  {Nishioka}}, \bibinfo {author} {\bibfnamefont {S.}~\bibnamefont {Ryu}}, \
  and\ \bibinfo {author} {\bibfnamefont {T.}~\bibnamefont {Takayanagi}},\
  }\href {\doibase 10.1088/1751-8113/42/50/504008} {\bibfield  {journal}
  {\bibinfo  {journal} {J. Phys. A}\ }\textbf {\bibinfo {volume} {42}},\
  \bibinfo {pages} {504008} (\bibinfo {year} {2009})},\ \Eprint
  {http://arxiv.org/abs/0905.0932} {arXiv:0905.0932 [hep-th]} \BibitemShut
  {NoStop}%
\bibitem [{\citenamefont {Rangamani}\ and\ \citenamefont
  {Takayanagi}(2017)}]{Rangamani:2016dms}%
  \BibitemOpen
  \bibfield  {author} {\bibinfo {author} {\bibfnamefont {M.}~\bibnamefont
  {Rangamani}}\ and\ \bibinfo {author} {\bibfnamefont {T.}~\bibnamefont
  {Takayanagi}},\ }\href {\doibase 10.1007/978-3-319-52573-0} {\emph {\bibinfo
  {title} {{Holographic Entanglement Entropy}}}},\ Vol.\ \bibinfo {volume}
  {931}\ (\bibinfo  {publisher} {Springer},\ \bibinfo {year} {2017})\ \Eprint
  {http://arxiv.org/abs/1609.01287} {arXiv:1609.01287 [hep-th]} \BibitemShut
  {NoStop}%
\bibitem [{\citenamefont {Chen}(2019)}]{Chen:2019lcd}%
  \BibitemOpen
  \bibfield  {author} {\bibinfo {author} {\bibfnamefont {B.}~\bibnamefont
  {Chen}},\ }\href {\doibase 10.1088/0253-6102/71/7/837} {\bibfield  {journal}
  {\bibinfo  {journal} {Commun. Theor. Phys.}\ }\textbf {\bibinfo {volume}
  {71}},\ \bibinfo {pages} {837} (\bibinfo {year} {2019})}\BibitemShut
  {NoStop}%
\bibitem [{\citenamefont {Nishioka}\ and\ \citenamefont
  {Takayanagi}(2007)}]{Nishioka:2006gr}%
  \BibitemOpen
  \bibfield  {author} {\bibinfo {author} {\bibfnamefont {T.}~\bibnamefont
  {Nishioka}}\ and\ \bibinfo {author} {\bibfnamefont {T.}~\bibnamefont
  {Takayanagi}},\ }\href {\doibase 10.1088/1126-6708/2007/01/090} {\bibfield
  {journal} {\bibinfo  {journal} {JHEP}\ }\textbf {\bibinfo {volume} {01}},\
  \bibinfo {pages} {090} (\bibinfo {year} {2007})},\ \Eprint
  {http://arxiv.org/abs/hep-th/0611035} {arXiv:hep-th/0611035} \BibitemShut
  {NoStop}%
\bibitem [{\citenamefont {Klebanov}\ \emph {et~al.}(2008)\citenamefont
  {Klebanov}, \citenamefont {Kutasov},\ and\ \citenamefont
  {Murugan}}]{Klebanov:2007ws}%
  \BibitemOpen
  \bibfield  {author} {\bibinfo {author} {\bibfnamefont {I.~R.}\ \bibnamefont
  {Klebanov}}, \bibinfo {author} {\bibfnamefont {D.}~\bibnamefont {Kutasov}}, \
  and\ \bibinfo {author} {\bibfnamefont {A.}~\bibnamefont {Murugan}},\ }\href
  {\doibase 10.1016/j.nuclphysb.2007.12.017} {\bibfield  {journal} {\bibinfo
  {journal} {Nucl. Phys. B}\ }\textbf {\bibinfo {volume} {796}},\ \bibinfo
  {pages} {274} (\bibinfo {year} {2008})},\ \Eprint
  {http://arxiv.org/abs/0709.2140} {arXiv:0709.2140 [hep-th]} \BibitemShut
  {NoStop}%
\bibitem [{\citenamefont {Bah}\ \emph {et~al.}(2009)\citenamefont {Bah},
  \citenamefont {Faraggi}, \citenamefont {Pando~Zayas},\ and\ \citenamefont
  {Terrero-Escalante}}]{Bah:2007kcs}%
  \BibitemOpen
  \bibfield  {author} {\bibinfo {author} {\bibfnamefont {I.}~\bibnamefont
  {Bah}}, \bibinfo {author} {\bibfnamefont {A.}~\bibnamefont {Faraggi}},
  \bibinfo {author} {\bibfnamefont {L.~A.}\ \bibnamefont {Pando~Zayas}}, \ and\
  \bibinfo {author} {\bibfnamefont {C.~A.}\ \bibnamefont {Terrero-Escalante}},\
  }\href {\doibase 10.1142/S0217751X0904542X} {\bibfield  {journal} {\bibinfo
  {journal} {Int. J. Mod. Phys. A}\ }\textbf {\bibinfo {volume} {24}},\
  \bibinfo {pages} {2703} (\bibinfo {year} {2009})},\ \Eprint
  {http://arxiv.org/abs/0710.5483} {arXiv:0710.5483 [hep-th]} \BibitemShut
  {NoStop}%
\bibitem [{\citenamefont {Bah}\ \emph {et~al.}(2012)\citenamefont {Bah},
  \citenamefont {Pando~Zayas},\ and\ \citenamefont
  {Terrero-Escalante}}]{Bah:2008cj}%
  \BibitemOpen
  \bibfield  {author} {\bibinfo {author} {\bibfnamefont {I.}~\bibnamefont
  {Bah}}, \bibinfo {author} {\bibfnamefont {L.~A.}\ \bibnamefont
  {Pando~Zayas}}, \ and\ \bibinfo {author} {\bibfnamefont {C.~A.}\ \bibnamefont
  {Terrero-Escalante}},\ }\href {\doibase 10.1142/S0217751X12500480} {\bibfield
   {journal} {\bibinfo  {journal} {Int. J. Mod. Phys. A}\ }\textbf {\bibinfo
  {volume} {27}},\ \bibinfo {pages} {1250048} (\bibinfo {year} {2012})},\
  \Eprint {http://arxiv.org/abs/0809.2912} {arXiv:0809.2912 [hep-th]}
  \BibitemShut {NoStop}%
\bibitem [{\citenamefont {Ali-Akbari}\ and\ \citenamefont
  {Lezgi}(2017)}]{Ali-Akbari:2017vtb}%
  \BibitemOpen
  \bibfield  {author} {\bibinfo {author} {\bibfnamefont {M.}~\bibnamefont
  {Ali-Akbari}}\ and\ \bibinfo {author} {\bibfnamefont {M.}~\bibnamefont
  {Lezgi}},\ }\href {\doibase 10.1103/PhysRevD.96.086014} {\bibfield  {journal}
  {\bibinfo  {journal} {Phys. Rev. D}\ }\textbf {\bibinfo {volume} {96}},\
  \bibinfo {pages} {086014} (\bibinfo {year} {2017})},\ \Eprint
  {http://arxiv.org/abs/1706.04335} {arXiv:1706.04335 [hep-th]} \BibitemShut
  {NoStop}%
\bibitem [{\citenamefont {Anber}\ and\ \citenamefont
  {Kolligs}(2018)}]{Anber:2018ohz}%
  \BibitemOpen
  \bibfield  {author} {\bibinfo {author} {\bibfnamefont {M.~M.}\ \bibnamefont
  {Anber}}\ and\ \bibinfo {author} {\bibfnamefont {B.~J.}\ \bibnamefont
  {Kolligs}},\ }\href {\doibase 10.1007/JHEP08(2018)175} {\bibfield  {journal}
  {\bibinfo  {journal} {JHEP}\ }\textbf {\bibinfo {volume} {08}},\ \bibinfo
  {pages} {175} (\bibinfo {year} {2018})},\ \Eprint
  {http://arxiv.org/abs/1804.01956} {arXiv:1804.01956 [hep-th]} \BibitemShut
  {NoStop}%
\bibitem [{\citenamefont {Jain}\ \emph {et~al.}(2023)\citenamefont {Jain},
  \citenamefont {Jena},\ and\ \citenamefont {Mahapatra}}]{Jain:2022hxl}%
  \BibitemOpen
  \bibfield  {author} {\bibinfo {author} {\bibfnamefont {P.}~\bibnamefont
  {Jain}}, \bibinfo {author} {\bibfnamefont {S.~S.}\ \bibnamefont {Jena}}, \
  and\ \bibinfo {author} {\bibfnamefont {S.}~\bibnamefont {Mahapatra}},\ }\href
  {\doibase 10.1103/PhysRevD.107.086016} {\bibfield  {journal} {\bibinfo
  {journal} {Phys. Rev. D}\ }\textbf {\bibinfo {volume} {107}},\ \bibinfo
  {pages} {086016} (\bibinfo {year} {2023})},\ \Eprint
  {http://arxiv.org/abs/2209.15355} {arXiv:2209.15355 [hep-th]} \BibitemShut
  {NoStop}%
\bibitem [{\citenamefont {da~Rocha}(2022)}]{daRocha:2021xwq}%
  \BibitemOpen
  \bibfield  {author} {\bibinfo {author} {\bibfnamefont {R.}~\bibnamefont
  {da~Rocha}},\ }\href {\doibase 10.1103/PhysRevD.105.026014} {\bibfield
  {journal} {\bibinfo  {journal} {Phys. Rev. D}\ }\textbf {\bibinfo {volume}
  {105}},\ \bibinfo {pages} {026014} (\bibinfo {year} {2022})},\ \Eprint
  {http://arxiv.org/abs/2111.01244} {arXiv:2111.01244 [hep-th]} \BibitemShut
  {NoStop}%
\bibitem [{\citenamefont {Zhang}\ and\ \citenamefont
  {Zhu}(2019)}]{Zhang:2019zbf}%
  \BibitemOpen
  \bibfield  {author} {\bibinfo {author} {\bibfnamefont {Z.-q.}\ \bibnamefont
  {Zhang}}\ and\ \bibinfo {author} {\bibfnamefont {X.}~\bibnamefont {Zhu}},\
  }\href {\doibase 10.1140/epja/i2019-12687-4} {\bibfield  {journal} {\bibinfo
  {journal} {Eur. Phys. J. A}\ }\textbf {\bibinfo {volume} {55}},\ \bibinfo
  {pages} {18} (\bibinfo {year} {2019})}\BibitemShut {NoStop}%
\bibitem [{\citenamefont {Rahimi}\ \emph {et~al.}(2017)\citenamefont {Rahimi},
  \citenamefont {Ali-Akbari},\ and\ \citenamefont {Lezgi}}]{Rahimi:2016bbv}%
  \BibitemOpen
  \bibfield  {author} {\bibinfo {author} {\bibfnamefont {M.}~\bibnamefont
  {Rahimi}}, \bibinfo {author} {\bibfnamefont {M.}~\bibnamefont {Ali-Akbari}},
  \ and\ \bibinfo {author} {\bibfnamefont {M.}~\bibnamefont {Lezgi}},\ }\href
  {\doibase 10.1016/j.physletb.2017.05.055} {\bibfield  {journal} {\bibinfo
  {journal} {Phys. Lett. B}\ }\textbf {\bibinfo {volume} {771}},\ \bibinfo
  {pages} {583} (\bibinfo {year} {2017})},\ \Eprint
  {http://arxiv.org/abs/1610.01835} {arXiv:1610.01835 [hep-th]} \BibitemShut
  {NoStop}%
\bibitem [{\citenamefont {Baggioli}\ \emph {et~al.}(2018)\citenamefont
  {Baggioli}, \citenamefont {Padhi}, \citenamefont {Phillips},\ and\
  \citenamefont {Setty}}]{Baggioli:2018afg}%
  \BibitemOpen
  \bibfield  {author} {\bibinfo {author} {\bibfnamefont {M.}~\bibnamefont
  {Baggioli}}, \bibinfo {author} {\bibfnamefont {B.}~\bibnamefont {Padhi}},
  \bibinfo {author} {\bibfnamefont {P.~W.}\ \bibnamefont {Phillips}}, \ and\
  \bibinfo {author} {\bibfnamefont {C.}~\bibnamefont {Setty}},\ }\href
  {\doibase 10.1007/JHEP07(2018)049} {\bibfield  {journal} {\bibinfo  {journal}
  {JHEP}\ }\textbf {\bibinfo {volume} {07}},\ \bibinfo {pages} {049} (\bibinfo
  {year} {2018})},\ \Eprint {http://arxiv.org/abs/1805.01470} {arXiv:1805.01470
  [hep-th]} \BibitemShut {NoStop}%
\bibitem [{\citenamefont {Baggioli}\ and\ \citenamefont
  {Giataganas}(2021)}]{Baggioli:2020cld}%
  \BibitemOpen
  \bibfield  {author} {\bibinfo {author} {\bibfnamefont {M.}~\bibnamefont
  {Baggioli}}\ and\ \bibinfo {author} {\bibfnamefont {D.}~\bibnamefont
  {Giataganas}},\ }\href {\doibase 10.1103/PhysRevD.103.026009} {\bibfield
  {journal} {\bibinfo  {journal} {Phys. Rev. D}\ }\textbf {\bibinfo {volume}
  {103}},\ \bibinfo {pages} {026009} (\bibinfo {year} {2021})},\ \Eprint
  {http://arxiv.org/abs/2007.07273} {arXiv:2007.07273 [hep-th]} \BibitemShut
  {NoStop}%
\bibitem [{\citenamefont {Baggioli}\ \emph {et~al.}(2023)\citenamefont
  {Baggioli}, \citenamefont {Liu},\ and\ \citenamefont
  {Wu}}]{Baggioli:2023ynu}%
  \BibitemOpen
  \bibfield  {author} {\bibinfo {author} {\bibfnamefont {M.}~\bibnamefont
  {Baggioli}}, \bibinfo {author} {\bibfnamefont {Y.}~\bibnamefont {Liu}}, \
  and\ \bibinfo {author} {\bibfnamefont {X.-M.}\ \bibnamefont {Wu}},\ }\href
  {\doibase 10.1007/JHEP05(2023)221} {\bibfield  {journal} {\bibinfo  {journal}
  {JHEP}\ }\textbf {\bibinfo {volume} {05}},\ \bibinfo {pages} {221} (\bibinfo
  {year} {2023})},\ \Eprint {http://arxiv.org/abs/2302.11096} {arXiv:2302.11096
  [hep-th]} \BibitemShut {NoStop}%
\bibitem [{\citenamefont {Dudal}\ and\ \citenamefont
  {Mahapatra}(2018)}]{Dudal:2018ztm}%
  \BibitemOpen
  \bibfield  {author} {\bibinfo {author} {\bibfnamefont {D.}~\bibnamefont
  {Dudal}}\ and\ \bibinfo {author} {\bibfnamefont {S.}~\bibnamefont
  {Mahapatra}},\ }\href {\doibase 10.1007/JHEP07(2018)120} {\bibfield
  {journal} {\bibinfo  {journal} {JHEP}\ }\textbf {\bibinfo {volume} {07}},\
  \bibinfo {pages} {120} (\bibinfo {year} {2018})},\ \Eprint
  {http://arxiv.org/abs/1805.02938} {arXiv:1805.02938 [hep-th]} \BibitemShut
  {NoStop}%
\bibitem [{\citenamefont {Mahapatra}(2019)}]{Mahapatra:2019uql}%
  \BibitemOpen
  \bibfield  {author} {\bibinfo {author} {\bibfnamefont {S.}~\bibnamefont
  {Mahapatra}},\ }\href {\doibase 10.1007/JHEP04(2019)137} {\bibfield
  {journal} {\bibinfo  {journal} {JHEP}\ }\textbf {\bibinfo {volume} {04}},\
  \bibinfo {pages} {137} (\bibinfo {year} {2019})},\ \Eprint
  {http://arxiv.org/abs/1903.05927} {arXiv:1903.05927 [hep-th]} \BibitemShut
  {NoStop}%
\bibitem [{\citenamefont {Aref'eva}(2019)}]{Arefeva:2019dvl}%
  \BibitemOpen
  \bibfield  {author} {\bibinfo {author} {\bibfnamefont {I.~Y.}\ \bibnamefont
  {Aref'eva}},\ }\href {\doibase 10.1134/S1547477119050042} {\bibfield
  {journal} {\bibinfo  {journal} {Phys. Part. Nucl. Lett.}\ }\textbf {\bibinfo
  {volume} {16}},\ \bibinfo {pages} {486} (\bibinfo {year} {2019})}\BibitemShut
  {NoStop}%
\bibitem [{\citenamefont {Knaute}\ and\ \citenamefont
  {K\"ampfer}(2017)}]{Knaute:2017lll}%
  \BibitemOpen
  \bibfield  {author} {\bibinfo {author} {\bibfnamefont {J.}~\bibnamefont
  {Knaute}}\ and\ \bibinfo {author} {\bibfnamefont {B.}~\bibnamefont
  {K\"ampfer}},\ }\href {\doibase 10.1103/PhysRevD.96.106003} {\bibfield
  {journal} {\bibinfo  {journal} {Phys. Rev. D}\ }\textbf {\bibinfo {volume}
  {96}},\ \bibinfo {pages} {106003} (\bibinfo {year} {2017})},\ \Eprint
  {http://arxiv.org/abs/1706.02647} {arXiv:1706.02647 [hep-ph]} \BibitemShut
  {NoStop}%
\bibitem [{\citenamefont {Li}\ \emph {et~al.}(2021)\citenamefont {Li},
  \citenamefont {Xu},\ and\ \citenamefont {Huang}}]{Li:2020pgn}%
  \BibitemOpen
  \bibfield  {author} {\bibinfo {author} {\bibfnamefont {Z.}~\bibnamefont
  {Li}}, \bibinfo {author} {\bibfnamefont {K.}~\bibnamefont {Xu}}, \ and\
  \bibinfo {author} {\bibfnamefont {M.}~\bibnamefont {Huang}},\ }\href
  {\doibase 10.1088/1674-1137/abc539} {\bibfield  {journal} {\bibinfo
  {journal} {Chin. Phys. C}\ }\textbf {\bibinfo {volume} {45}},\ \bibinfo
  {pages} {013116} (\bibinfo {year} {2021})},\ \Eprint
  {http://arxiv.org/abs/2002.08650} {arXiv:2002.08650 [hep-th]} \BibitemShut
  {NoStop}%
\bibitem [{\citenamefont {Aref'eva}\ \emph {et~al.}(2020)\citenamefont
  {Aref'eva}, \citenamefont {Patrushev},\ and\ \citenamefont
  {Slepov}}]{Arefeva:2020uec}%
  \BibitemOpen
  \bibfield  {author} {\bibinfo {author} {\bibfnamefont {I.~Y.}\ \bibnamefont
  {Aref'eva}}, \bibinfo {author} {\bibfnamefont {A.}~\bibnamefont {Patrushev}},
  \ and\ \bibinfo {author} {\bibfnamefont {P.}~\bibnamefont {Slepov}},\ }\href
  {\doibase 10.1007/JHEP07(2020)043} {\bibfield  {journal} {\bibinfo  {journal}
  {JHEP}\ }\textbf {\bibinfo {volume} {07}},\ \bibinfo {pages} {043} (\bibinfo
  {year} {2020})},\ \Eprint {http://arxiv.org/abs/2003.05847} {arXiv:2003.05847
  [hep-th]} \BibitemShut {NoStop}%
\bibitem [{\citenamefont {Asadi}\ \emph {et~al.}(2023)\citenamefont {Asadi},
  \citenamefont {Amrahi},\ and\ \citenamefont
  {Eshaghi-Kenari}}]{Asadi:2022mvo}%
  \BibitemOpen
  \bibfield  {author} {\bibinfo {author} {\bibfnamefont {M.}~\bibnamefont
  {Asadi}}, \bibinfo {author} {\bibfnamefont {B.}~\bibnamefont {Amrahi}}, \
  and\ \bibinfo {author} {\bibfnamefont {H.}~\bibnamefont {Eshaghi-Kenari}},\
  }\href {\doibase 10.1140/epjc/s10052-023-11214-6} {\bibfield  {journal}
  {\bibinfo  {journal} {Eur. Phys. J. C}\ }\textbf {\bibinfo {volume} {83}},\
  \bibinfo {pages} {69} (\bibinfo {year} {2023})},\ \Eprint
  {http://arxiv.org/abs/2209.01586} {arXiv:2209.01586 [hep-th]} \BibitemShut
  {NoStop}%
\bibitem [{\citenamefont {Jokela}\ and\ \citenamefont
  {Subils}(2021)}]{Jokela:2020wgs}%
  \BibitemOpen
  \bibfield  {author} {\bibinfo {author} {\bibfnamefont {N.}~\bibnamefont
  {Jokela}}\ and\ \bibinfo {author} {\bibfnamefont {J.~G.}\ \bibnamefont
  {Subils}},\ }\href {\doibase 10.1007/JHEP02(2021)147} {\bibfield  {journal}
  {\bibinfo  {journal} {JHEP}\ }\textbf {\bibinfo {volume} {02}},\ \bibinfo
  {pages} {147} (\bibinfo {year} {2021})},\ \Eprint
  {http://arxiv.org/abs/2010.09392} {arXiv:2010.09392 [hep-th]} \BibitemShut
  {NoStop}%
\bibitem [{\citenamefont {Fujita}\ \emph {et~al.}(2020)\citenamefont {Fujita},
  \citenamefont {He},\ and\ \citenamefont {Sun}}]{Fujita:2020qvp}%
  \BibitemOpen
  \bibfield  {author} {\bibinfo {author} {\bibfnamefont {M.}~\bibnamefont
  {Fujita}}, \bibinfo {author} {\bibfnamefont {S.}~\bibnamefont {He}}, \ and\
  \bibinfo {author} {\bibfnamefont {Y.}~\bibnamefont {Sun}},\ }\href {\doibase
  10.1103/PhysRevD.102.126019} {\bibfield  {journal} {\bibinfo  {journal}
  {Phys. Rev. D}\ }\textbf {\bibinfo {volume} {102}},\ \bibinfo {pages}
  {126019} (\bibinfo {year} {2020})},\ \Eprint
  {http://arxiv.org/abs/2005.01048} {arXiv:2005.01048 [hep-th]} \BibitemShut
  {NoStop}%
\bibitem [{\citenamefont {Li}(2024)}]{Li:2024lrh}%
  \BibitemOpen
  \bibfield  {author} {\bibinfo {author} {\bibfnamefont {Z.}~\bibnamefont
  {Li}},\ }\href {\doibase 10.1103/PhysRevD.110.046012} {\bibfield  {journal}
  {\bibinfo  {journal} {Phys. Rev. D}\ }\textbf {\bibinfo {volume} {110}},\
  \bibinfo {pages} {046012} (\bibinfo {year} {2024})},\ \Eprint
  {http://arxiv.org/abs/2402.02944} {arXiv:2402.02944 [hep-th]} \BibitemShut
  {NoStop}%
\bibitem [{\citenamefont {Buividovich}\ and\ \citenamefont
  {Polikarpov}(2008)}]{Buividovich:2008kq}%
  \BibitemOpen
  \bibfield  {author} {\bibinfo {author} {\bibfnamefont {P.~V.}\ \bibnamefont
  {Buividovich}}\ and\ \bibinfo {author} {\bibfnamefont {M.~I.}\ \bibnamefont
  {Polikarpov}},\ }\href {\doibase 10.1016/j.nuclphysb.2008.04.024} {\bibfield
  {journal} {\bibinfo  {journal} {Nucl. Phys. B}\ }\textbf {\bibinfo {volume}
  {802}},\ \bibinfo {pages} {458} (\bibinfo {year} {2008})},\ \Eprint
  {http://arxiv.org/abs/0802.4247} {arXiv:0802.4247 [hep-lat]} \BibitemShut
  {NoStop}%
\bibitem [{\citenamefont {Itou}\ \emph {et~al.}(2016)\citenamefont {Itou},
  \citenamefont {Nagata}, \citenamefont {Nakagawa}, \citenamefont {Nakamura},\
  and\ \citenamefont {Zakharov}}]{Itou:2015cyu}%
  \BibitemOpen
  \bibfield  {author} {\bibinfo {author} {\bibfnamefont {E.}~\bibnamefont
  {Itou}}, \bibinfo {author} {\bibfnamefont {K.}~\bibnamefont {Nagata}},
  \bibinfo {author} {\bibfnamefont {Y.}~\bibnamefont {Nakagawa}}, \bibinfo
  {author} {\bibfnamefont {A.}~\bibnamefont {Nakamura}}, \ and\ \bibinfo
  {author} {\bibfnamefont {V.~I.}\ \bibnamefont {Zakharov}},\ }\href {\doibase
  10.1093/ptep/ptw050} {\bibfield  {journal} {\bibinfo  {journal} {PTEP}\
  }\textbf {\bibinfo {volume} {2016}},\ \bibinfo {pages} {061B01} (\bibinfo
  {year} {2016})},\ \Eprint {http://arxiv.org/abs/1512.01334} {arXiv:1512.01334
  [hep-th]} \BibitemShut {NoStop}%
\bibitem [{\citenamefont {Klebanov}\ and\ \citenamefont
  {Witten}(1999)}]{Klebanov:1999tb}%
  \BibitemOpen
  \bibfield  {author} {\bibinfo {author} {\bibfnamefont {I.~R.}\ \bibnamefont
  {Klebanov}}\ and\ \bibinfo {author} {\bibfnamefont {E.}~\bibnamefont
  {Witten}},\ }\href {\doibase 10.1016/S0550-3213(99)00387-9} {\bibfield
  {journal} {\bibinfo  {journal} {Nucl. Phys. B}\ }\textbf {\bibinfo {volume}
  {556}},\ \bibinfo {pages} {89} (\bibinfo {year} {1999})},\ \Eprint
  {http://arxiv.org/abs/hep-th/9905104} {arXiv:hep-th/9905104} \BibitemShut
  {NoStop}%
\bibitem [{\citenamefont {Cherman}\ \emph {et~al.}(2009)\citenamefont
  {Cherman}, \citenamefont {Cohen},\ and\ \citenamefont
  {Werbos}}]{Cherman:2008eh}%
  \BibitemOpen
  \bibfield  {author} {\bibinfo {author} {\bibfnamefont {A.}~\bibnamefont
  {Cherman}}, \bibinfo {author} {\bibfnamefont {T.~D.}\ \bibnamefont {Cohen}},
  \ and\ \bibinfo {author} {\bibfnamefont {E.~S.}\ \bibnamefont {Werbos}},\
  }\href {\doibase 10.1103/PhysRevC.79.045203} {\bibfield  {journal} {\bibinfo
  {journal} {Phys. Rev. C}\ }\textbf {\bibinfo {volume} {79}},\ \bibinfo
  {pages} {045203} (\bibinfo {year} {2009})},\ \Eprint
  {http://arxiv.org/abs/0804.1096} {arXiv:0804.1096 [hep-ph]} \BibitemShut
  {NoStop}%
\bibitem [{\citenamefont {Afonin}(2011)}]{Afonin:2010hn}%
  \BibitemOpen
  \bibfield  {author} {\bibinfo {author} {\bibfnamefont {S.~S.}\ \bibnamefont
  {Afonin}},\ }\href {\doibase 10.1142/S0217751X11053997} {\bibfield  {journal}
  {\bibinfo  {journal} {Int. J. Mod. Phys. A}\ }\textbf {\bibinfo {volume}
  {26}},\ \bibinfo {pages} {3615} (\bibinfo {year} {2011})},\ \Eprint
  {http://arxiv.org/abs/1012.5065} {arXiv:1012.5065 [hep-ph]} \BibitemShut
  {NoStop}%
\bibitem [{\citenamefont {Casini}\ and\ \citenamefont
  {Huerta}(2004)}]{Casini:2004bw}%
  \BibitemOpen
  \bibfield  {author} {\bibinfo {author} {\bibfnamefont {H.}~\bibnamefont
  {Casini}}\ and\ \bibinfo {author} {\bibfnamefont {M.}~\bibnamefont
  {Huerta}},\ }\href {\doibase 10.1016/j.physletb.2004.08.072} {\bibfield
  {journal} {\bibinfo  {journal} {Phys. Lett. B}\ }\textbf {\bibinfo {volume}
  {600}},\ \bibinfo {pages} {142} (\bibinfo {year} {2004})},\ \Eprint
  {http://arxiv.org/abs/hep-th/0405111} {arXiv:hep-th/0405111} \BibitemShut
  {NoStop}%
\bibitem [{\citenamefont {Casini}\ and\ \citenamefont
  {Huerta}(2007)}]{Casini:2006es}%
  \BibitemOpen
  \bibfield  {author} {\bibinfo {author} {\bibfnamefont {H.}~\bibnamefont
  {Casini}}\ and\ \bibinfo {author} {\bibfnamefont {M.}~\bibnamefont
  {Huerta}},\ }\href {\doibase 10.1088/1751-8113/40/25/S57} {\bibfield
  {journal} {\bibinfo  {journal} {J. Phys. A}\ }\textbf {\bibinfo {volume}
  {40}},\ \bibinfo {pages} {7031} (\bibinfo {year} {2007})},\ \Eprint
  {http://arxiv.org/abs/cond-mat/0610375} {arXiv:cond-mat/0610375} \BibitemShut
  {NoStop}%
\bibitem [{\citenamefont {Liu}\ and\ \citenamefont
  {Mezei}(2013)}]{Liu:2012eea}%
  \BibitemOpen
  \bibfield  {author} {\bibinfo {author} {\bibfnamefont {H.}~\bibnamefont
  {Liu}}\ and\ \bibinfo {author} {\bibfnamefont {M.}~\bibnamefont {Mezei}},\
  }\href {\doibase 10.1007/JHEP04(2013)162} {\bibfield  {journal} {\bibinfo
  {journal} {JHEP}\ }\textbf {\bibinfo {volume} {04}},\ \bibinfo {pages} {162}
  (\bibinfo {year} {2013})},\ \Eprint {http://arxiv.org/abs/1202.2070}
  {arXiv:1202.2070 [hep-th]} \BibitemShut {NoStop}%
\end{thebibliography}%

\end{document}